\documentclass[nonacm, sigconf]{acmart}

\usepackage{booktabs}
\usepackage{multirow}
\usepackage{graphicx}
\usepackage{subcaption}
\usepackage{algorithm}
\usepackage{algpseudocode}
\usepackage{tikz}
\usepackage{xspace}
\usepackage{xcolor}
\usepackage{soul}
\usepackage{tcolorbox}

\newtcolorbox{mybasecolorbox}[1][]{%
 colback=gray!10, colframe=black,
 coltitle=black, fonttitle=\small\bfseries,
 sharp corners,
 width=1.0\linewidth,
 title=#1,
 left=1pt,
 right=1pt, 
 top=1pt, 
 bottom=1pt, 
 boxsep=2pt, 
 boxrule=0.5pt,
 before=\vspace{1pt}, 
 after=\vspace{1pt},
 }

\newenvironment{mytitlebox}[1][]{%
 \centering
 \mybasecolorbox[#1]
 \linespread{0.98}\selectfont
 \itshape
}{%
 \endmybasecolorbox
}

\newenvironment{myinlinebox}[1][]{%
 \mytitlebox
 {\upshape\bfseries #1}%
}{%
 \endmybasecolorbox
}

\AtBeginDocument{%
}

\definecolor{darkgreen}{rgb}{0.0, 0.5, 0.0}

\newcommand{\fig}[1]{Figure~\ref{#1}}
\newcommand{\sect}[1]{Section~\ref{#1}}
\newcommand{\tab}[1]{Table~\ref{#1}}
\newcommand{\algo}[1]{Algorithm~\ref{#1}}

\fancypagestyle{firstpageheader}{
    \fancyhf{} 
    \fancyhead[C]{Preprint} 
    
}

\begin{document}

\title{Debunking the CUDA Myth Towards GPU-based AI Systems
\\ \Large : Evaluation of the Performance and Programmability of Intel's Gaudi NPU for AI Model Serving
}

\makeatletter
\def\@authornotemark{} 
\def\@fnsymbol#1{} 
\makeatother

\author{
\centering
\makebox[2.5cm][c]{Yunjae Lee\textsuperscript{$\dagger\S$}}\hspace{1.5em}\makebox[2.5cm][c]{Juntaek Lim\textsuperscript{$\dagger\S$}}\\[0.5\baselineskip]
\makebox[2.5cm][c]{Jehyeon Bang\textsuperscript{$\dagger$}}\hspace{1.5em}\makebox[2.5cm][c]{Eunyeong Cho\textsuperscript{$\dagger$}}\hspace{1.5em}\makebox[2.5cm][c]{Huijong Jeong\textsuperscript{$\ddagger$}}\hspace{1.5em}\makebox[2.5cm][c]{Taesu Kim\textsuperscript{$\ddagger$}}\hspace{1.5em}\makebox[2.5cm][c]{Hyungjun Kim\textsuperscript{$\ddagger$}}\\[0.5\baselineskip]
\makebox[2.5cm][c]{Joonhyung Lee\textsuperscript{$*$}}\hspace{1.5em}\makebox[2.5cm][c]{Jinseop Im\textsuperscript{$\dagger$}}\hspace{1.5em}\makebox[2.5cm][c]{Ranggi Hwang\textsuperscript{$\dagger$}}\hspace{1.5em}\makebox[2.5cm][c]{Se Jung Kwon\textsuperscript{$*$}}\hspace{1.5em}\makebox[2.5cm][c]{Dongsoo Lee\textsuperscript{$*$}}\\[0.5\baselineskip]
Minsoo Rhu\textsuperscript{$\dagger\diamond$}\\[\baselineskip]
\makebox[3cm][c]{\textsuperscript{$\dagger$}KAIST}
\makebox[3cm][c]{\textsuperscript{$*$}NAVER Cloud}
\makebox[3cm][c]{\textsuperscript{$\ddagger$}SqueezeBits}
}

\authornote{\textsuperscript{$\S$}Co-first authors who contributed equally to this research.}
\authornote{\textsuperscript{$\diamond$}Corresponding author: Minsoo Rhu (\href{mailto:mrhu@kaist.ac.kr}{\textcolor{blue}{mrhu@kaist.ac.kr}})}

\begin{abstract}

This paper presents a comprehensive evaluation of Intel Gaudi NPUs as an alternative to NVIDIA GPUs, which is currently the de facto standard in AI system design. First, we create a suite of microbenchmarks to compare Intel Gaudi-2 with NVIDIA A100, showing that Gaudi-2 achieves competitive performance not only in primitive AI compute, memory, and communication operations but also in executing several important AI workloads end-to-end. We then assess Gaudi NPU’s programmability by discussing several software-level optimization strategies to employ for implementing critical FBGEMM operators and vLLM, evaluating their efficiency against GPU-optimized counterparts. Results indicate that Gaudi-2 achieves energy efficiency comparable to A100, though there are notable areas for improvement in terms of software maturity.  Overall, we conclude that, with effective integration into high-level AI frameworks, Gaudi NPUs could challenge NVIDIA GPU's dominance in the AI server market, though further improvements are necessary to fully compete with NVIDIA’s robust software ecosystem.

\end{abstract}
\maketitle
\thispagestyle{plain}
\pagestyle{plain}
\setcounter{page}{1}
\section{Introduction} 
\label{sect:intro}

\begin{quotation}
\noindent \em
``In the past few years there have been many studies claiming GPUs deliver substantial speedups (10$\times$$-$1,000$\times$) over multi-core CPUs on throughput computing kernels. $\ldots$ 
After applying optimizations appropriate for both CPUs and GPUs the performance gap between NVIDIA GTX280 and Intel Core i7 960 narrows to only 2.5$\times$'' 

  \emph{``Debunking the 100X GPU vs. CPU Myth: An Evaluation of Throughput Computing on CPU and GPU''}, Intel, ISCA, 2010~\cite{debunking_100x_gpu_cpu_myth}
\end{quotation}
\vspace{0.5em}

Intel's latency-optimized processor architectures have dominated the computing industry for decades, serving as the foundation for executing a wide range of  applications. 
However, as GPU computing gained prominence in the late 2000s, NVIDIA, then the underdog, began to challenge Intel's long-standing dominance in the server market. With the rise of AI, throughput-optimized processor architectures spearheaded by NVIDIA GPUs dethroned Intel, establishing NVIDIA's CUDA software ecosystem as the de facto standard for training and deploying AI models. 

One might argue that domain-specific architectures for AI, also known as Neural Processing Units (NPUs), present competitive alternatives to GPUs. However, cases of successfully utilizing NPUs for deploying AI services are limited to a handful of hyperscalers that can amortize the enormous development costs of NPUs by serving millions to billions of customers with their AI offerings (e.g., Google's TPU~\cite{google_tpuv4}, Meta's MTIA~\cite{mtia_isca2023}, Amazon's Inferentia~\cite{inferentia}).
As a result, most AI services deployed today are built using NVIDIA GPUs. The primary advantage of NVIDIA GPUs over commercially available NPUs is their ease of programming with CUDA. The flexible programming interface of CUDA, along with the rich software ecosystem built around GPU-accelerated backend libraries (e.g., cuBLAS, cuDNN, cuSPARSE, cuSOLVER, cuDF, cuVS~\cite{cublas,cudnn,cusparse,cusolver,cudf,cuvs}), allows developers to easily implement and optimize new AI models created by AI practitioners (e.g., state-space models like Mamba~\cite{mamba_official_github}). While some cloud service providers like Google and Amazon do offer NPUs for developers in the form of ``AI-as-a-Service''~\cite{aws_machine_learning,google_vertex_ai}, these platforms provide only limited access and programmability for the backend NPUs, such as Google TPU~\cite{google_tpu_website} and Amazon Inferentia~\cite{inferentia}. This limitation makes it challenging to perform low-level kernel implementation and performance optimization specific to the target backend NPU architecture. 

Given this landscape, Intel's Gaudi NPU~\cite{gaudi2_white_paper} is noteworthy for several reasons. First,  Gaudi NPUs come with a native programming language called TPC-C (the CUDA equivalent for Gaudi)~\cite{gaudi_sdk_tpc_user_guide} as well as low-level compute primitives that ease the implementation of compute kernels targeting the NPU's compute engines. Second, the performance of end-to-end AI applications utilizing these compute kernels is (according to Intel's claims) comparable to, and in some cases better than, that of NVIDIA GPUs. Third, Gaudi NPUs are currently widely available for purchase, allowing researchers to thoroughly characterize this new NPU device vs. NVIDIA GPUs.

To this end, this paper presents a detailed characterization of Intel's Gaudi NPU for AI model serving, assessing whether Intel, now the underdog, can pose a tangible threat to NVIDIA's seemingly unassailable dominance in the AI computing market. 
A thorough understanding of this emerging NPU architecture and its applicability to various AI workloads can offer valuable insights for programmers, AI service providers, and computer architects working on next-generation NPU designs. As such, we conduct a comprehensive analysis of the Gaudi NPU from multiple dimensions, evaluating not just its raw performance but also its programmability for facilitating performance optimization and AI model development.

\begin{table}[]
  \caption{Comparison of NVIDIA A100 and Intel Gaudi-2.}
  \vspace{-1em}
  \centering
  \resizebox{\columnwidth}{!}{%
  \begin{tabular}{cc|cc|c}
  \hline
  \multicolumn{2}{c|}{} & \multicolumn{1}{c|}{\textbf{NVIDIA A100}} & \textbf{Intel Gaudi-2} & \textbf{Ratio} \\ \hline
  \multicolumn{1}{c|}{\multirow{2}{*}{\textbf{Compute}}} & \multirow{2}{*}{TFLOPS (BF16)} & \multicolumn{1}{c|}{312 (Tensor Cores)} & 432 (MME) & 1.4$\times$ \\ \cline{3-5} 
  \multicolumn{1}{c|}{} &  & \multicolumn{1}{c|}{39 (SIMD Cores)} & 11 (TPC) & 0.3$\times$ \\ \hline
  \multicolumn{1}{c|}{\multirow{4}{*}{\textbf{Memory}}} & HBM Type & \multicolumn{2}{c|}{HBM2E} & - \\ \cline{2-5} 
  \multicolumn{1}{c|}{} & HBM Capacity & \multicolumn{1}{c|}{80 GB} & 96 GB & 1.2$\times$ \\ \cline{2-5} 
  \multicolumn{1}{c|}{} & HBM bandwidth & \multicolumn{1}{c|}{2 TB/sec} & 2.46 TB/sec & 1.2$\times$ \\ \cline{2-5} 
  \multicolumn{1}{c|}{} & SRAM capacity & \multicolumn{1}{c|}{40 MB (L2 Cache)} & 48 MB (Shared) & 1.2$\times$ \\ \hline
  \multicolumn{2}{c|}{\textbf{Communication}} & \multicolumn{1}{c|}{\begin{tabular}[c]{@{}c@{}}600 GB/sec bidirectional\\ (NVLink)\end{tabular}} & \begin{tabular}[c]{@{}c@{}}600 GB/sec bidirectional\\ (RoCE)\end{tabular} & 1.0$\times$ \\ \hline
  \multicolumn{2}{c|}{\textbf{Power}} & \multicolumn{1}{c|}{400 Watts} & 600 Watts & 1.5$\times$ \\ \hline
  \end{tabular}%
  }
  \label{tab:gpu_vs_hpu}
  \vspace{-.5em}
  \end{table}
  
{\bf (Performance)}
To enable detailed experimental studies and analyses, we first develop a set of microbenchmarks targeting Gaudi NPUs to stress-test their ability to maximize performance in several key compute, memory, and communication primitives. 
In this work, we use Intel's second-generation Gaudi NPU (Gaudi-2) and NVIDIA’s A100 GPU as comparison points\footnote{The hardware and software architecture of Intel's recently announced Gaudi-3 is virtually identical to that of Gaudi-2 (but with limited availability), except that Gaudi-3 offers higher compute and memory throughput, thanks to its chiplet-based design.}, as both processors are manufactured using TSMC’s 7nm technology node and are supported by an HBM2E-based memory subsystem, providing comparable performance (\tab{tab:gpu_vs_hpu}).
Our microbenchmarking revealed that Gaudi-2 demonstrates highly competitive performance in important primitive AI operations, particularly for tasks involving regular compute and memory accesses. However, Gaudi-2 did fall short of A100 in certain scenarios involving fine-grained data accesses and collective communications across a small number of processors.

After confirming Gaudi NPU's competitiveness against NVIDIA GPUs in primitive AI operations through our microbenchmark-based analysis, we next evaluate both systems at the end-to-end AI application level. Specifically, we focus on recommendation systems (RecSys) and large language models (LLMs), as these two are among the most widely deployed AI models in cloud environments. Our analysis of end-to-end AI applications revealed that Gaudi-2 achieves 28\% lower energy-efficiency for RecSys but 50\% higher energy-efficiency for LLMs compared to the A100.

{\bf (Programmability)} We also present case studies on utilizing the Gaudi NPU’s programming interface to optimize its performance. Specifically, we discuss software-level optimizations that can be employed to develop Gaudi NPU-optimized versions of FBGEMM's Batched Embedding Table~\cite{fbgemm_batchedtable} and vLLM~\cite{vllm}, which enable high-performance model serving for RecSys and LLMs, respectively. Initially, publicly available Gaudi-optimized software for RecSys embedding layers and vLLM showed underwhelming results, achieving only 37\% and 6\% of the performance seen in their GPU-optimized counterparts. However, through various software-level optimizations applied at the low-level TPC-C (Batched Embedding Table) and high-level PyTorch (vLLM), we show that the performance-optimized Gaudi-2 can achieve 80\% and 101\% of A100's performance running end-to-end RecSys and LLM applications based on the state-of-the-art FBGEMM (RecSys) and vLLM (LLM).

Overall, we conclude that the Gaudi NPU has significant potential to emerge as a contender to NVIDIA GPUs for AI model serving, challenging NVIDIA's dominance in the AI computing industry. 
Most AI practitioners use high-level AI frameworks like PyTorch or TensorFlow for model development. As long as NPU chip vendors effectively support these frameworks with optimized low-level backend libraries, our analysis suggests that NVIDIA’s CUDA programming system might not be as formidable a ``moat'' in the AI server market. In other words, our conclusion is that the strength of NVIDIA GPU-based AI systems lies in its rich software ecosystem, rather than in CUDA itself.
However, it is important not to misconstrue our assessment as an overly optimistic outlook on Intel's Gaudi NPUs. NVIDIA's stronghold in AI still remains robust, and we believe that Gaudi NPUs still face several key challenges that should be addressed to effectively compete with NVIDIA's established position, which we further discuss in \sect{sect:discussion}.

To summarize our {\bf key contributions}:
  
\begin{itemize}
\item To the best of our knowledge, this work is the first to provide a detailed characterization of Intel's Gaudi NPUs compared to NVIDIA GPUs, examining not only their performance but, more crucially, their programmability.
\item We implement a set of microbenchmarks to conduct a comparative study with NVIDIA GPUs, analyzing the potential, limitations, and bottlenecks of the Gaudi NPU architecture.
\item To assess Gaudi NPU's programmability, we discuss important software-level optimization strategies to employ to develop Gaudi-optimized versions of DLRM~\cite{facebook_dlrm} and vLLM~\cite{vllm}. Through this exercise, we discuss the strengths and weaknesses of the Gaudi NPU's software architecture.

\end{itemize}

\section{Background}
\label{sect:background}

\subsection{Intel Gaudi Hardware Architecture}
\label{sect:gaudi_npu_hw_arch}

\begin{figure}[t]
\centering
\includegraphics[width=0.485\textwidth]{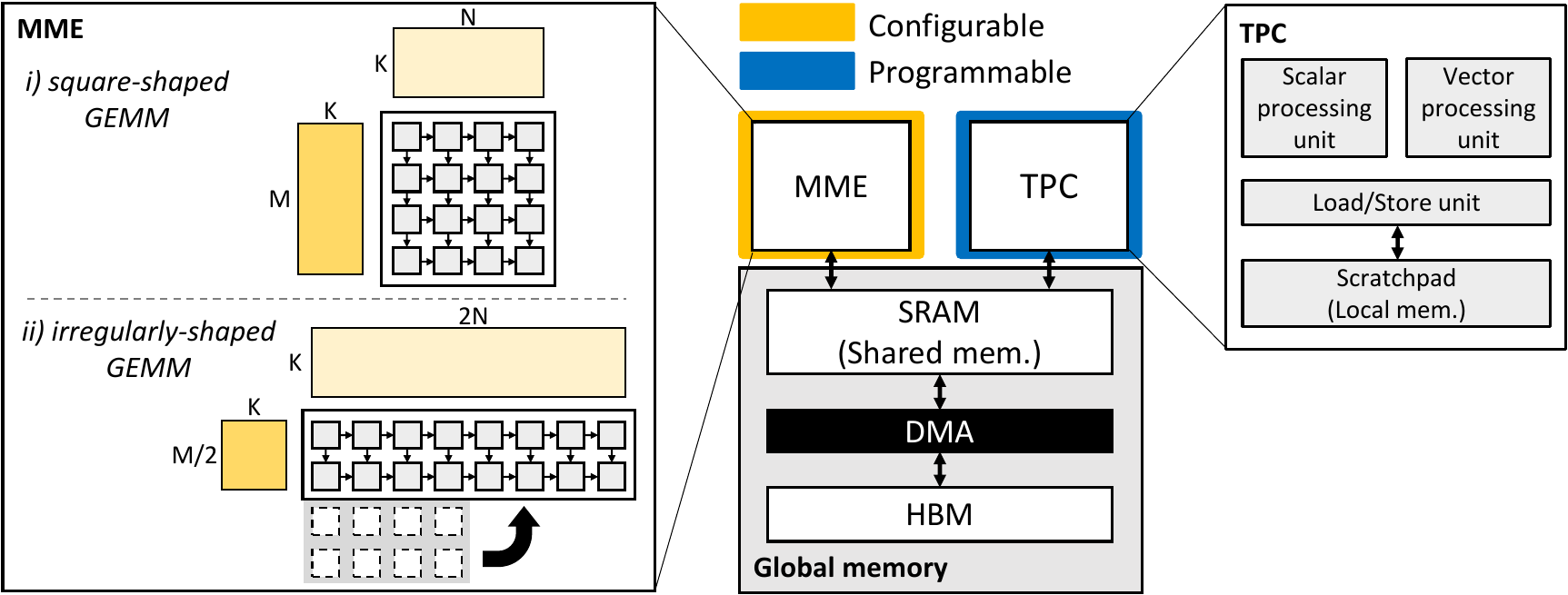} 
\vspace{-1.3em}
\caption{High-level overview of Intel's Gaudi NPU architecture.}
\label{fig:gaudi_arch}
\vspace{-1.3em}
\end{figure}

{\bf (Compute)} 
The Gaudi processor architecture is designed based on a \emph{heterogeneous} compute paradigm, integrating two key components (\fig{fig:gaudi_arch}): Matrix Multiplication Engines (MMEs) and fully programmable Tensor Processing Cores (TPCs). Gaudi-2 features two MMEs and 24 TPCs, which together provide high throughput by pipelining computations between the MMEs and TPCs. 

The MME is a large, output-stationary systolic array with a 256$\times$256 MAC (Multiply-Accumulate) structure~\cite{hotchips_gaudi}, designed to handle general matrix multiplication (GEMM) workloads, such as fully connected, convolutional, and batched GEMM layers. The MME is designed to be highly \emph{configurable} in order to maximize the utilization of its MAC array. Specifically, the two MMEs in Gaudi-2, originally composed of two separate 256$\times$256 MAC units, can be dynamically reconfigured at runtime as a single 512$\times$256 MAC unit, a single 1024$\times$128 MAC unit, and others, depending on the shape of the input and output matrices of the GEMM operation. The optimal MME configuration for each target GEMM  is determined by the Gaudi graph compiler, which we discuss in \sect{sect:benchmarking_compute}. Similar to NVIDIA's Tensor Cores~\cite{tensor_cores}, the MME is a co-processor purposefully designed to accelerate matrix multiplications. As such, the MME is not directly programmable, meaning users cannot alter its functionality and can only utilize it for matrix multiplications. 

Unlike the MME, the TPC is a highly programmable, VLIW (Very Long Instruction Word)-based processor designed to execute multiple types of instructions in parallel. Each instruction type is processed by dedicated units that handle load/store operations and scalar/vector operations (\fig{fig:gaudi_arch}), enabling efficient parallel execution. The SIMD (Single Instruction, Multiple Data) vector unit can handle 2048-bit wide vector operations. This makes the TPC highly effective for various data-parallel tasks in AI, particularly for nonlinear and non-matrix-based computations, such as vector gather-scatter operations or activation functions.

In terms of performance, the MME in Gaudi-2 delivers up to 432 TFLOPS of throughput for BF16 (brain floating point 16-bit~\cite{bfloat16}) operations. The TPCs provide an additional 11 TFLOPS for BF16. In comparison, NVIDIA's A100 offers 312 TFLOPS for matrix operations (using Tensor Cores) and 39 TFLOPS for vector operations (using SIMD Cores). In total, Gaudi-2 delivers approximately 1.26 times in aggregate higher compute throughput than A100 (\tab{tab:gpu_vs_hpu}).

{\bf (Memory)} Gaudi-2 features 96 GB of HBM2E, delivering a bandwidth of 2.45 TB/sec (\tab{tab:gpu_vs_hpu}). High memory bandwidth plays a crucial role in AI workloads, particularly in memory-bound tasks such as the embedding vector gathers in RecSys~\cite{facebook_dlrm, dlrm_dcnv2}
and the decoding stages of LLMs\cite{chatgpt, palm, llama}. The A100 offers 2 TB/sec of bandwidth, making Gaudi-2 approximately $20\%$ higher in terms of maximum memory throughput. Regarding on-chip storage, Gaudi-2 includes 48 MB of on-chip SRAM, referred to as shared memory, which serves as a scratchpad for the Gaudi graph compiler. This shared memory acts as temporary storage, facilitating data movement between the MMEs, TPCs, and DMA engines to maximize both on-chip data reuse and hardware utilization.  Each TPC has its own local memory (used as a scratchpad within the TPC), divided into scalar and vector memory banks. The scalar memory in Gaudi-2 TPC is 1 KB in size and is accessed in 4-byte aligned chunks, while the vector memory is 80 KB and is accessed in 128- or 256-byte chunks. These local memories are private to each TPC, ensuring fast, dedicated memory operations without interference from other TPCs. In contrast, global memory (including the on-chip shared memory and off-chip HBM), is accessible to the entire system with a minimum access granularity of 256-byte chunks.

{\bf (Communication)} 
Intel's HLS-Gaudi-2 server~\cite{hls_gaudi2} is integrated with eight Gaudi-2 chips. Each Gaudi-2 is equipped with 24$\times$100 GbE RoCEv2~\cite{roce} ports, providing a maximum bandwidth of 2.4 Tbps when all eight chips participate in collective communication. Of the 24 RoCE ports, 21 are dedicated to direct, point-to-point (P2P) inter-chip communication, with each pair of Gaudi-2 chips connected by three 100 GbE links. Since any given pair of Gaudi-2 is connected via P2P links, the effective bandwidth depends on the number of Gaudi-2 chips involved in the collective communication. For example, when two Gaudi-2 chips communicate, only 300 Gbps of communication bandwidth is available (3$\times$100 GbE links), which is just 1/8 of the maximum 2.4 Tbps bandwidth. In contrast, NVIDIA's DGX A100 server is integrated with a network switch (NVSwitch~\cite{nvswitch}) that enables all GPUs within the node to communicate simultaneously at the total NVLink bandwidth. Unlike P2P connections where multiple processors must split bandwidth, NVSwitch ensures that each GPU can transfer data at maximum speed, regardless of how many GPUs are involved in communication. Consequently, for AI model serving that do not fully utilize all eight Gaudi-2 chips, such dynamic scaling can result in limited communication bandwidth, affecting system-wide performance.

\subsection{Intel Gaudi Software Architecture}
\label{subsec:gaudi_npu_sw_stack}

\begin{figure*}[t]
\centering
\includegraphics[width=0.75\textwidth]{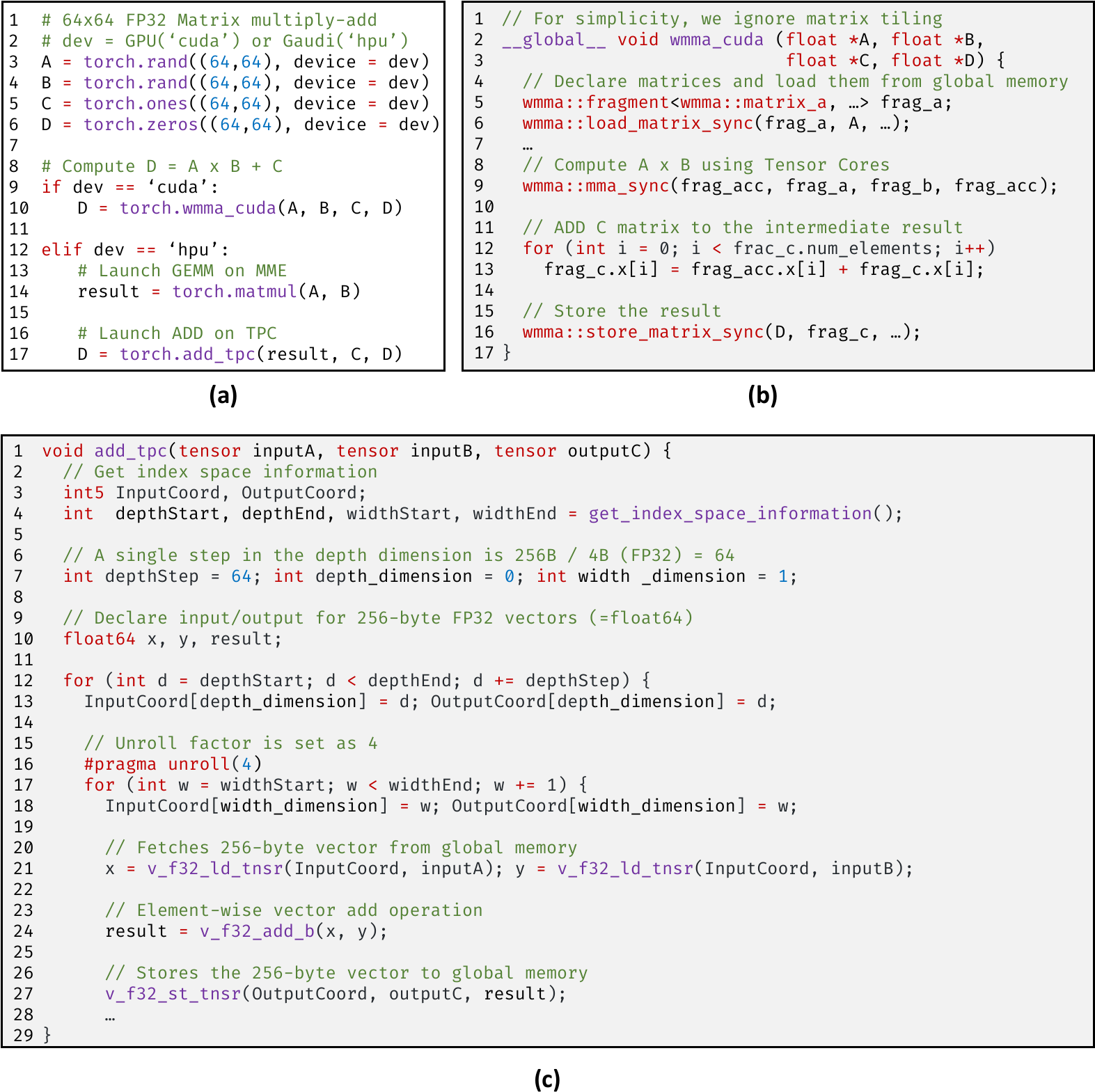} 
\vspace{-0.5em}
\caption{Example code showing how a matrix multiply-add operation can be programmed for execution on an NVIDIA GPU and Intel Gaudi.  (a) At the PyTorch level, the corresponding low-level kernels are executed based on the device type (i.e., GPU (``\texttt{cuda}'', line 9) or Gaudi (``\texttt{hpu}'', line 12)). (b) In NVIDIA's CUDA, the matrix multiply-add operation is performed within a single kernel launch using the \texttt{WMMA} APIs, which leverage Tensor Cores alongside SIMD Cores.  (c) In Gaudi, the GEMM operation can only be handled at the PyTorch level (line 14 in (a)), so a TPC-C kernel (\texttt{add\_tpc}) is called at the PyTorch level to execute the subsequent add operation (line 17 in (a)).}
\label{fig:example_code}
\vspace{-1em}
\end{figure*}

{\bf (Programming model)} 
The GPU programming system follows the Single Instruction Multiple Thread (SIMT) model, which falls under the Single Program Multiple Data (SPMD) paradigm. To support efficient SIMT execution, GPUs include unique microarchitectural support, such as a large register file for fine-grained, massive multithreading~\mbox{\cite{gebhart:isca:2011}}, dynamic branch divergence resolution~\mbox{\cite{dwf:fung:micro,rhu:2013:dpe,sho:arxiv:cfm}}, and warp-wide memory coalescing~\mbox{\cite{chatterjee:memory_divergence,meng:memory_divergence}}. In contrast, Gaudi lacks these features and instead uses a \emph{single-threaded} programming model optimized for data-level rather than thread-level parallelism.

In NVIDIA's CUDA, programmers can leverage the \texttt{WMMA} (Warp Matrix Multiply and Accumulate) APIs~\mbox{\cite{wmma}} to directly utilize Tensor Cores alongside conventional SIMD Cores for computation within low-level CUDA kernels. Furthermore, these operations can be extended to a higher-level PyTorch API (line 10 of \mbox{\fig{fig:example_code}}(a) and line 9 of \mbox{\fig{fig:example_code}}(b)). In contrast, the Gaudi SDK currently restricts direct access to the MME units, allowing programmers to explicitly control only the TPCs. Instead, access to MME units is limited to the PyTorch level (\mbox{\fig{fig:example_code}}(a), line 14), thereby constraining programmers from directly optimizing performance involving both MME and TPC at a lower level. This limitation can be addressed through the Gaudi graph compiler's optimization pass, which we will discuss later. In Gaudi's TPC programming model, the target workload is partitioned across different TPCs, which execute the same TPC program. Workload distribution is performed by partitioning the \emph{index space} (equivalent to CUDA \emph{grid}), enabling each TPC to process different data independently. The index space can be divided up to five dimensions, and each member of the index space is allocated with an indivisible unit of work processed by a single TPC. 

\begin{figure}[t]
\centering
\vspace{0.3em}
\includegraphics[width=0.485\textwidth]{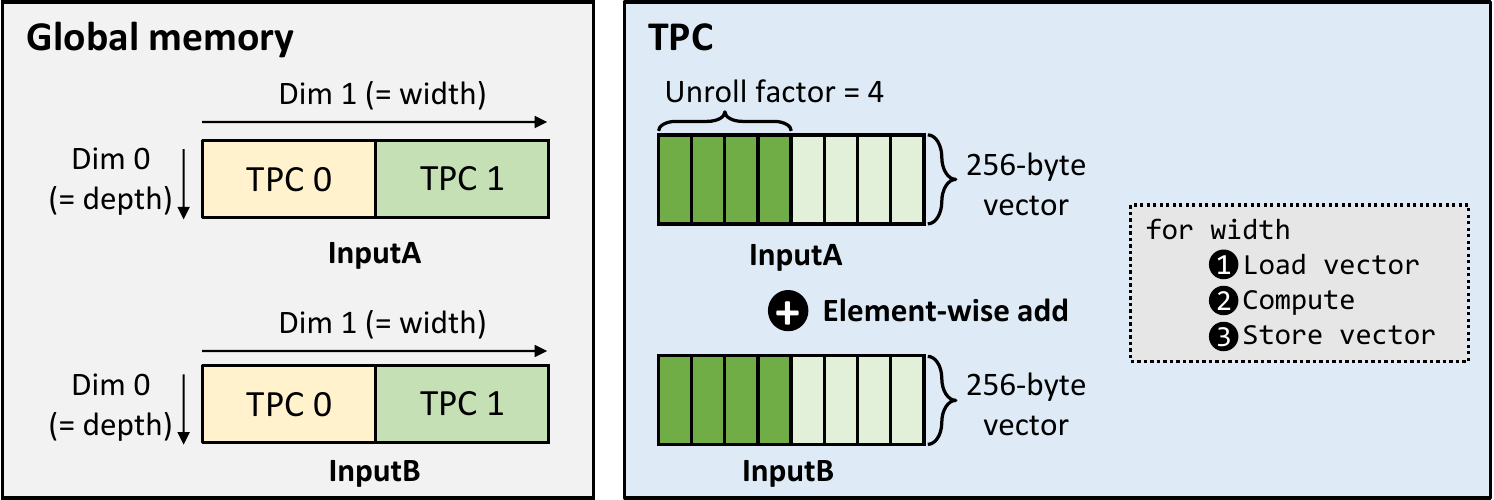} 
\vspace{-1.3em}
\caption{High-level overview of the TPC programming model. Example assumes a program performing an element-wise vector addition operation, partitioned into two dimensions of index space, with each partition being executed by a single TPC. The loop within the TPC program is assumed to be unrolled by a factor of ``4'' to maximize both instruction-level parallelism and memory-level parallelism.
}
\label{fig:tpc_programming_model}
\vspace{-1.3em}
\end{figure}

\fig{fig:tpc_programming_model} illustrates how a single TPC operates to conduct a simple element-wise vector addition operation (pseudo-code shown in \fig{fig:example_code}(c)). A TPC program is typically structured using a for-loop, which iteratively executes \emph{vector-wide} ``Load$\rightarrow$Compute$\rightarrow$Store'' operations. For optimal performance, the TPC programmers are advised to carefully partition the index space across the TPCs while maximizing the performance of individual TPCs. Two important best practices are recommended for TPC programmers. First, to maximize memory bandwidth utilization, the TPC's data access granularity should be aligned to 256 bytes as this is the minimum access granularity for global memory (e.g., line 7 in \mbox{\fig{fig:example_code}}(c) and a single step in the \texttt{depth} dimension in \fig{fig:tpc_programming_model}, which determines the memory access granularity, is sized at 256-bytes). Second, to fully utilize the TPC processor, it is recommended that programmers manually \emph{unroll} the for-loop to maximize both instruction-level and memory-level parallelism (e.g., a single step in the \texttt{width} dimension, which determines the magnitude of loop unrolling and thus the number of parallel operations, is sized at 4). This recommendation stems from the fact that TPC instructions have an average architectural latency of 4 processor cycles (i.e., the effect of executing a TPC instruction is reflected in the architectural state 4 cycles later)~\cite{habana_manual}. By unrolling four TPC instructions within a given iteration of a for-loop, the processor pipeline can be better utilized and memory access latency can be better hidden. To reduce the burden of manual loop unrolling, the TPC programming system provides preprocessor directives such as ``\texttt{\#pragma unroll}'' (line 16 in \mbox{\fig{fig:example_code}}(c)).

{\bf (Graph compiler)} Intel's Gaudi NPU comes with a software suite called Intel Gaudi SDK~\cite{gaudi_sdk}, which is tightly integrated with PyTorch and TensorFlow. This integration allows developers to utilize familiar tools while taking advantage of Gaudi NPU's hardware acceleration. Gaudi SDK includes a \emph{graph compiler} that converts AI models into a format optimized for execution on Gaudi NPUs. The graph compiler applies high-level model optimizations, such as operator and kernel fusion. For example, an MLIR~\mbox{\cite{lattner:mlir}}-based operation fuser selects arbitrary subgraphs of element-wise, reduction, and normalization operations, then JIT-fuses and compiles them into TPC kernels~\mbox{\cite{gaudi_mlir, hotchips_gaudi, gaudi3_white_paper}}. This improves performance by reducing memory bandwidth usage and enabling tensor shape-aware optimizations, unlike shape-agnostic kernel libraries. The graph compiler also performs hardware-specific optimizations, which include MME configuration adjustments for high MME utilization (\fig{fig:gaudi_arch}, further discussed in \sect{sect:benchmarking_compute}) and operator pipelining between MME and TPC. When an MME operation is followed by a TPC operation (e.g., GEMM followed by an activation function), the graph compiler breaks them into smaller, independent sub-operations to enable pipelined execution. This approach helps hide latency and reduce overall execution time. The graph compiler controls the process of orchestrating data transfers between the MME and TPC for pipelined execution, using on-chip shared memory as an intermediate buffer to minimize redundant off-chip memory accesses.

While lowering the target model graph into Gaudi NPU executable operations is crucial for performance optimization, the programmer, unfortunately, has no control over the graph compiler's optimization process. In other words, users cannot modify the behavior of the graph compiler nor dictate when a particular graph compiler optimization pass should be activated or not.

\section{Characterizing Gaudi NPU Performance}
\subsection{Motivation and Evaluation Methodology}
\label{subsect:evaluation_methodology}

\begin{table}[]
\caption{Evaluated microbenchmarks.}
  \vspace{-1em}
\centering
\renewcommand{\arraystretch}{1.1} 
\large
\resizebox{\columnwidth}{!}{%
\begin{tabular}{c|c|c|c}
\hline
\multicolumn{2}{c|}{\bf{Microbenchmark}} & \bf{System} & \bf{Implementation}  \\ \hline
\multirow{4}{*}{Compute} & \multirow{2}{*}{GEMM} & Gaudi-2 & PyTorch API \\ \cline{3-4} 
 &  & A100 & PyTorch API \\ \cline{2-4} 
 & \multirow{2}{*}{non-GEMM} & Gaudi-2 & TPC-C \\ \cline{3-4} 
 &  & A100 & CUDA  \\ \hline
\multirow{2}{*}{Memory} & \multirow{2}{*}{Vector gather-scatter} & Gaudi-2 & TPC-C \\ \cline{3-4} 
 &  & A100 & CUDA \\ \hline
\multirow{2}{*}{Communication} & \multirow{2}{*}{Collective communication} & Gaudi-2 & Intel HCCL~\cite{hccl} \\ \cline{3-4} 
 &  & A100 & NVIDIA NCCL~\cite{nccl} \\ \hline
\end{tabular}%
\vspace{-2.0em}
}
\label{tab:microbenchmark_table}
\end{table}

\begin{table}[]
    \caption{Evaluated end-to-end AI workloads.}
  \vspace{-1em}
    \centering
    \large
    \resizebox{\columnwidth}{!}{%
    \begin{tabular}{c|c|l|l|l}
    \hline
    \multicolumn{2}{c|}{\bf{Model}} & \multicolumn{1}{c|}{\bf{Embedding layer}} & \multicolumn{1}{c|}{\bf{MLP layer}} & \multicolumn{1}{c}{\bf{Interaction layer}} \\ \hline
    \multirow{2}{*}[-1.5em]{\centering \begin{tabular}{@{}c@{}}DLRM-DCNv2\\\cite{mlperf_recsys_benchmark}\end{tabular}} & RM1 & \begin{tabular}[l]{@{}l@{}}\# tables: 10\\ \# embeddings: 1M \\ \# gathers: 10\end{tabular} & \begin{tabular}[l]{@{}l@{}}Bottom: 512-256-64\\ Top: 1024-1024-512-256-1\end{tabular} & \begin{tabular}[l]{@{}l@{}}Low rank dim: 512\\ \# layers: 3\end{tabular} \\ \cline{2-5}
    & RM2 & \begin{tabular}[l]{@{}l@{}}\# tables: 20\\ \# embeddings: 1M\\ \# gathers: 100\end{tabular} & \begin{tabular}[l]{@{}l@{}}Bottom: 256-64-64\\ Top: 128-64-1\end{tabular} & \begin{tabular}[l]{@{}l@{}}Low rank dim: 64\\ \# layers: 2\end{tabular} \\ \hline
    \multicolumn{5}{c}{}   \\ \hline
    \multicolumn{2}{c|}{\bf{Model}} & \multicolumn{1}{c|}{\bf{Embedding layer}} & \multicolumn{2}{c}{\bf{Decoder layer}} \\ \hline
    \multirow{2}{*}[-1.5em]{\centering \begin{tabular}{@{}c@{}}Llama-3.1\\\cite{llama3}\end{tabular}} & 8B & \# vocabularies: 128,256 & \multicolumn{2}{l}{\begin{tabular}[l]{@{}l@{}}\# layers: 32\\ \# heads for query: 32\\ \# heads for key, value: 8\\ hidden/intermediate size: 4,096/14,336\end{tabular}} \\ \cline{2-5}
    & 70B & \# vocabularies: 128,256 & \multicolumn{2}{l}{\begin{tabular}[l]{@{}l@{}}\# layers: 80\\ \# heads for query: 64\\ \# heads for key, value: 8\\ hidden/intermediate size: 8,192/28,672\end{tabular}} \\ \hline
    \end{tabular}%
    }
    \label{tab:AI_workloads_table}
\vspace{-0.5em}
\end{table}

{\bf (Motivation)} AI practitioners utilize high-level AI software frameworks like PyTorch for model development. Consequently, a competitive AI software ecosystem should provide not only highly optimized, low-level backend libraries that accelerate performance-critical primitive AI operations (e.g., GEMM, vector gather-scatter, collective communication), but, more critically, it should also deliver the performance benefits of hardware acceleration at the end-to-end AI application level. To this end, the primary objective of our characterization is twofold. First, we develop microbenchmarks using Intel Gaudi SDK and our custom-designed TPC-C kernels to evaluate Gaudi's ability to achieve high-performance in key primitive AI operations (\tab{tab:microbenchmark_table}). Second, we set out to explore whether Gaudi can deliver competitive performance at the end-to-end AI application level. In our end-to-end performance characterization, we use recommendation systems (RecSys) and large language models (LLMs), as they represent the two most widely deployed AI models in today's datacenters while exhibiting very different compute and memory characteristics (\tab{tab:AI_workloads_table}). 

{\bf (Methodology)} All experiments discussed in the rest of this paper are conducted using an HLS-Gaudi-2 server (which contains eight Gaudi-2 chips connected via RoCE) and a DGX A100 server (which contains eight A100 GPUs connected via NVSwitch and NVLink) (\tab{tab:gpu_vs_hpu}). On the software side, we use Intel Gaudi Software v1.18.0, which is based on PyTorch 2.4, along with the TPC-C SDK for custom Gaudi kernel development. For the GPU system evaluation, we use PyTorch 2.4 and CUDA 12.4 for GPU kernel development.
All experimental results presented in this paper assume the BF16 data type, except when we evaluate end-to-end RecSys models which utilize FP32. The non-RecSys evaluation results for FP32 and their key takeaways were practically identical to those for BF16, so we omit presenting those results for brevity. 

For our microbenchmark analysis targeting primitive AI operations (\sect{sect:benchmarking_compute} to \sect{sect:microbenchmark_comm}), we focus on absolute performance and its resource utilization. When evaluating GEMM operations, we used PyTorch API, which defaults to cuBLAS~\mbox{\cite{cublas}} as its backend library, choosing the most optimal GEMM implementation per target device. For end-to-end AI workload analysis (\sect{sect:end_to_end_analysis}), we evaluate both performance and energy-efficiency. When evaluating the energy-efficiency of end-to-end AI workloads, each system's power consumption is measured using \texttt{nvidia-smi}~\cite{nvidia_smi} for A100 and \texttt{hl-smi}~\cite{hl_smi} for Gaudi-2.

\label{subsec:microbenchmarks}

\begin{figure}[t]
\centering
\includegraphics[width=0.485\textwidth]{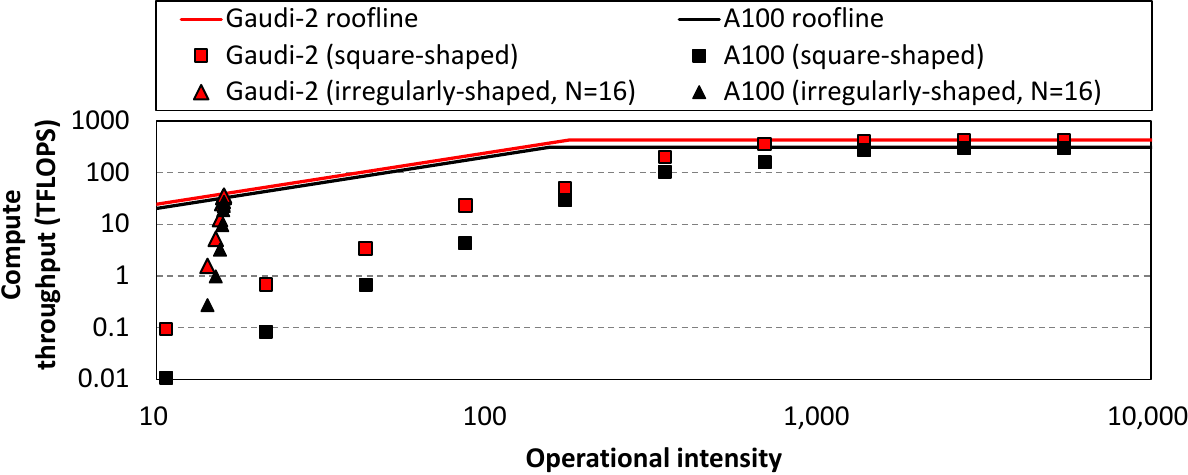} 
\vspace{-2em}
\caption{Roofline model showing the achieved TFLOPS of Gaudi-2 and A100 (BF16). The irregularly-shaped GEMMs, represented by triangle markers, have the $N$ dimension size fixed at 16. Performance is measured when both Gaudi-2 and A100 are configured to operate at its maximum possible frequency.}
\label{fig:mme_roofline}
\vspace{-1.3em}
\end{figure}

\subsection{Primivite ``Compute'' Operations}
\label{sect:benchmarking_compute}

\textbf{(MME for GEMM operations)} We first analyze the efficiency of Gaudi MMEs in conducting GEMM operations. GEMM involves multiplying two matrices, matrix $A$ of size ($M \times K$) and $B$ of size ($K \times N$), to produce a result matrix $C$ of size ($M \times N$). In \fig{fig:mme_roofline}, we compare the performance of various ($M$,$K$,$N$) GEMM shapes on Gaudi-2 and A100 by plotting the achieved TFLOPS using a roofline model. To simplify our discussion, we classify GEMM operations into two types: (1) \emph{square-shaped} GEMM (represented by square markers), where the dimensions $M$, $K$, and $N$ are all equal; and (2) \emph{irregularly-shaped} GEMM (represented by triangle markers), where dimension $N$ is set to a relatively small value compared to $M$ and $K$, resulting in input matrices $A$ and $B$ that are tall and skinny, exhibiting the properties of memory-bound GEMV operations. As shown in \fig{fig:mme_roofline}, Gaudi-2 consistently outperforms A100 across all ($M$,$K$,$N$) GEMM shapes we explore in this study. Notably, Gaudi-2 achieves 429 TFLOPS when $M$=$K$=$N$=$8192$, reaching $99.3\%$ of its  peak compute throughput (\tab{tab:gpu_vs_hpu}). Part of Gaudi-2’s higher absolute GEMM performance is due to its superior hardware specifications; its MME provides a maximum of 432 TFLOPS, which is $40\%$ higher than the A100’s 312 TFLOPS  offered by its Tensor Cores. Therefore, we also compare how \emph{efficiently} these two processors are able to utilize its hardware resources by measuring their compute ``utilization'' during GEMM executions. 

\begin{figure}[t]
\centering
\includegraphics[width=0.485\textwidth]{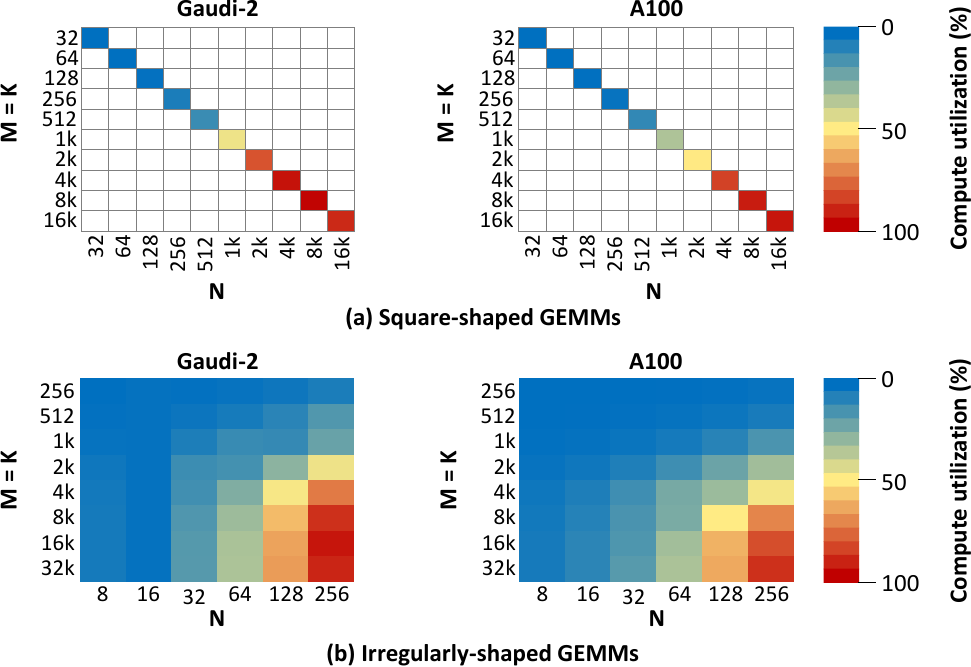} 
\caption{
Compute utilization when the GEMM operations are (a) square-shaped ($M$=$K$=$N$) and (b) irregularly-shaped ($M$ and $K$ are relatively larger than the fixed $N$). In all the thermal plots presented in this paper, warmer colors indicate higher values. In (a), we leave the GEMM shapes that do not satisfy $M$=$K$=$N$ vacant.
}
\label{fig:mme_square_n_gemv_util}
\end{figure}

\begin{figure}[t]
\centering
\includegraphics[width=0.485\textwidth]{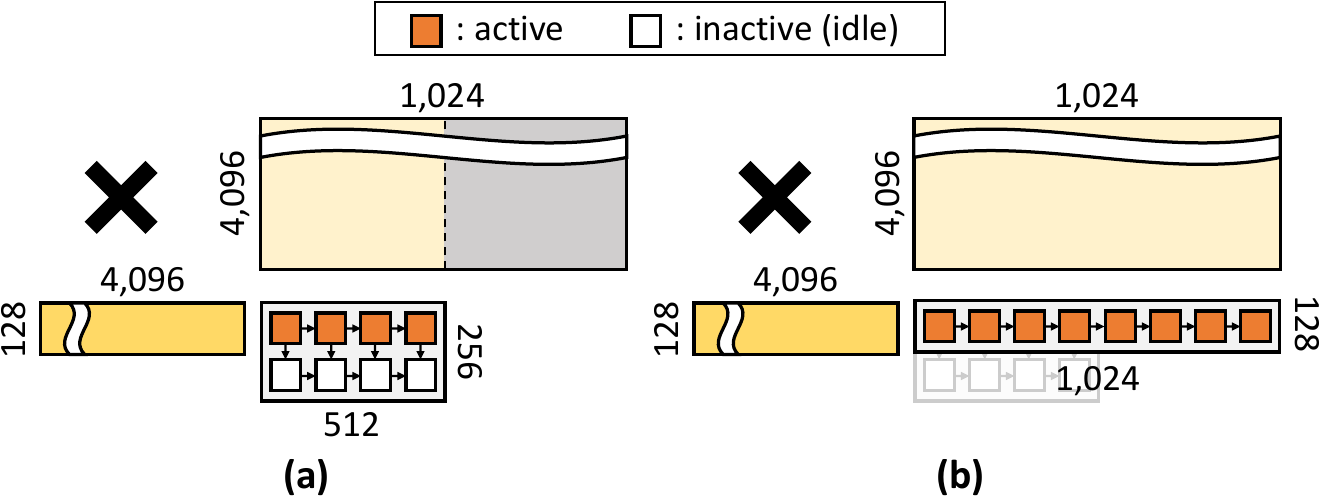}
\caption{Examples of how an irregularly-shaped GEMM is handled (a) by a typical output-stationary systolic array and (b) by Gaudi MME's systolic array \emph{with} reconfigurability.}
\label{fig:example_sa}
\vspace{-1em}
\end{figure}

\begin{figure}[t]
\centering
\includegraphics[width=0.485\textwidth]{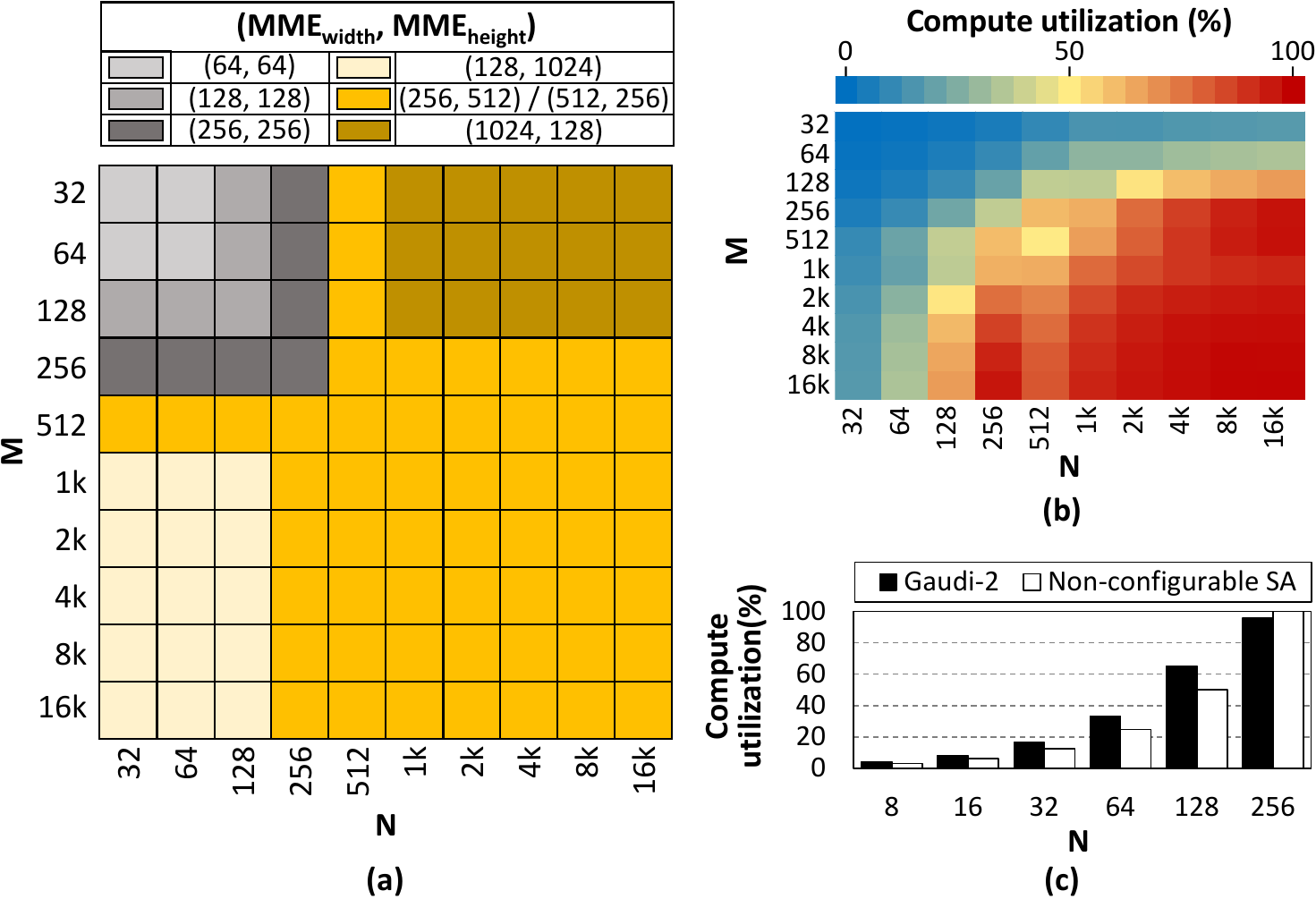}
\caption{(a) How the geometry of the MME systolic array (i.e., width (MME$_{width}$) and height (MME$_{height}$) of MME) is configured based on the ($M$, $N$) of GEMM while $K$ is fixed to 16,384, and (b) the corresponding compute utilization measured on Gaudi-2. To illustrate how MME's configurability enhances compute utilization, we compare (c) the \emph{measured} compute utilization when the MME   executes these GEMMs (black bars) vs. the theoretical, \emph{calculated} compute utilization when a non-configurable, output-stationary (256$\times$256$\times$2) systolic array with the same maximum FLOPS to Gaudi-2, as depicted in \mbox{\fig{fig:example_sa}(a)} (white bars), executes these GEMMs (results in (c) assume $M$=$K$=16,384, while varying $N$). It is worth pointing out that the gray-colored MME configuration in (a) activates only a \emph{subset} of the maximum (256$\times$256$\times$2) MAC array. Intel does not disclose specific details on the MME’s microarchitecture or its configurability; one possible explanation is that for smaller GEMM shapes, the MME power-gates the inactive portions of the MAC array to save energy.} 
\label{fig:mme_graph_compiler}
\vspace{-1.5em}
\end{figure}

In \fig{fig:mme_square_n_gemv_util}, we measure the ratio of \emph{achieved} TFLOPS to \emph{peak} TFLOPS to quantify the compute utilization of GEMM operations. The results indicate that Gaudi-2 not only achieves higher absolute TFLOPS (\fig{fig:mme_roofline}) but also outperforms A100 in terms of compute utilization. Across all evaluated data points, Gaudi-2 achieves an average $4.5\%$ higher compute utilization (maximum of $32\%$ when $M$=$K$=$N$=2,048) than A100. These results were counterintuitive to our initial expectations because large systolic arrays, as employed in Gaudi MMEs, are known to suffer from low MAC utilization when the GEMM operation is irregularly shaped and, therefore, not optimally aligned with the geometry of the systolic array~\cite{diva,tpu1,planaria}. 
For example, in a typical output-stationary systolic array, when the GEMM's $M$ and $N$ dimensions are smaller than the height and width of the systolic array, the MAC units can experience significant underutilization (\fig{fig:example_sa}(a)). As discussed in \sect{sect:background}, however, recall that Gaudi's MME can dynamically \emph{reconfigure} the geometry  of its systolic array (i.e., height and width dimensions) to better align with the target GEMM's ($M$,$K$,$N$) shape, significantly enhancing the utility of their MAC units (\fig{fig:example_sa}(b)). To better understand this behavior, we use the Intel Gaudi Profiler to reverse-engineer how the graph compiler and runtime system manages MME’s GEMM execution, which provide hints on how the MME geometry is dynamically configured in relation to the target ($M$,$K$,$N$) GEMM shape. In \fig{fig:mme_graph_compiler}, we summarize the results of our reverse-engineering, showing how the geometry of the MME systolic array is configured as a function of the input GEMM’s $M$ and $N$ dimension sizes while fixing $K$=$16,384$ (\fig{fig:mme_graph_compiler}(a)) and how this configuration translates into the MME’s compute utilization (\fig{fig:mme_graph_compiler}(b)). Compared to a typical output-stationary systolic array design \emph{without} reconfigurability (\fig{fig:mme_graph_compiler}(c)), the configurable MME architecture provides up to  $15\%$ improvement in compute utilization vs. non-configurable, output-stationary systolic array.

\begin{myinlinebox}[Key takeaway \#1:]
  When performing GEMM, Gaudi-2 achieved both higher absolute performance and greater compute utilization than A100. This superior GEMM performance and efficiency can be attributed not only to Gaudi-2's higher max compute throughput but, more importantly, to the configurability of the Gaudi-2 MME, which enables its systolic array to flexibly adapt its geometry to be most optimal for the target GEMM's ($M$,$K$,$N$) shape.
\end{myinlinebox}

{\bf (TPC for non-GEMM operations)} We now evaluate the performance of Gaudi-2's TPC in conducting vector operations, which is critical in performing AI operations like activation functions. Our microbenchmarks for non-GEMM operations are designed based on the STREAM~\cite{stream_benchmark} benchmark suite, which measures sustainable compute throughput and memory bandwidth for element-wise vector operations. \algo{algo:stream_benchmark} summarizes these three microbenchmarks: ADD, SCALE, and TRIAD. These microbenchmarks access two (SCALE) or three (ADD, TRIAD) arrays in a \emph{streaming} fashion. The number of floating-point operations involved is 1 for both ADD and SCALE (addition and multiplication, respectively) and 2 for TRIAD (multiplication followed by addition). We implement these microbenchmarks using TPC-C, where each TPC executes the ADD, SCALE, or TRIAD operation over a dedicated set of array elements (a total of 24 million elements) assigned to its specific index space (\fig{fig:tpc_programming_model}). We utilize these microbenchmarks to demonstrate how performance can be optimized by applying the two TPC programming best practices discussed in \sect{subsec:gaudi_npu_sw_stack}, namely (1) the need to align data access granularity in 256 bytes, and (2) the importance of unrolling loops to maximize parallelism.

\begin{algorithm}[t!]
  \caption{Microbenchmarks for evaluating non-GEMM operations}
  \label{algo:stream_benchmark}
 \begin{footnotesize}
\begin{algorithmic}[1]
\Procedure{Add}{$a$, $b$, $c$, $N$}
    \For{$i = 0$ \textbf{to} $N-1$}
        \State $c[i] \gets a[i] + b[i]$
    \EndFor
\EndProcedure
\\
\Procedure{Scale}{$a$, $b$, $scalar$, $N$}
    \For{$i = 0$ \textbf{to} $N-1$}
        \State $b[i] \gets scalar \times a[i]$
    \EndFor
\EndProcedure
\\
\Procedure{Triad}{$a$, $b$, $c$, $scalar$, $N$}
    \For{$i = 0$ \textbf{to} $N-1$}
        \State $c[i] \gets scalar \times a[i] + b[i]$
    \EndFor
\EndProcedure
\end{algorithmic}
\end{footnotesize}
\end{algorithm} 

\begin{figure}[t]
\centering
\vspace{-1em}
\includegraphics[width=0.49\textwidth]{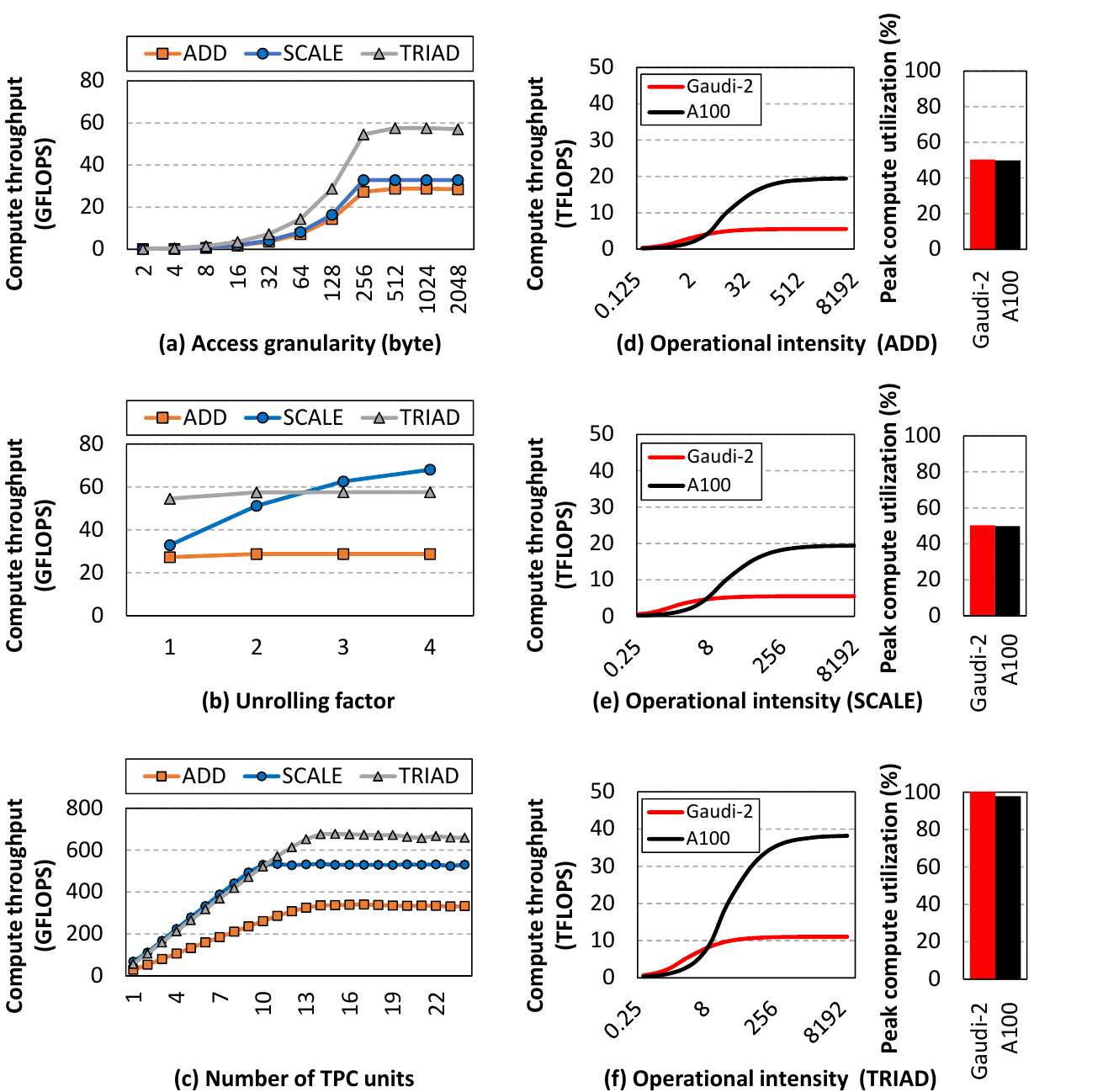} 
\caption{Compute throughput of ADD, TRIAD, and SCALE over a vector with 24 million scalar elements (BF16). The effects of (a) data access granularity and (b) unrolling factor on a single TPC's throughput are shown in figures (a and b). In (c), we scale out the number of TPCs by weak scaling the three microbenchmarks. The left axis on figures (d, e, and f) illustrate the compute throughput when the operational intensity of (d) ADD, (e) SCALE, and (f) TRIAD is artificially increased. The right axis on these figures (d, e, and f) shows the compute utilization when Gaudi-2 and A100's compute throughput peaks at its saturation point.
}
\label{fig:tpc_compute_throughput}
\vspace{-1.8em}
\end{figure}

We first show how a \emph{single} TPC's performance is improved by applying the two aforementioned best practices. In \fig{fig:tpc_compute_throughput}(a), we measure the compute throughput by varying our microbenchmarks' data access granularity from 2 to 2,048 bytes, \emph{without} loop unrolling. The results clearly show the significant performance drop when the data access granularity is set lower than 256 bytes, which is Gaudi's minimum memory access granularity. At data access granularities higher than 256 bytes, the overall throughput saturates at around 55 GFLOPS for TRIAD and around 30 GFLOPS for SCALE and ADD. From this point, we explore how much further performance improvements can be achieved by unrolling the for-loop in \algo{algo:stream_benchmark} (discussed in \sect{subsec:gaudi_npu_sw_stack}). As shown in \fig{fig:tpc_compute_throughput}(b), the compute throughput of SCALE improves remarkably, while ADD and TRIAD achieve only slight improvements as the loop unrolling factor increases. Both TRIAD and ADD load vectors from two arrays, resulting in two load instructions and one compute instruction (\texttt{v\_bf16\_mac\_b} for TRIAD and \texttt{v\_bf16\_add\_b} for ADD) for each loop iteration. In contrast, SCALE requires only one load instruction and one compute instruction (\texttt{v\_bf16\_mul\_b}) from a single array, for each loop iteration, providing more pipeline opportunities to hide the 4 TPC processor cycle latency and benefiting more from loop unrolling. Overall, these results highlight the importance of loop unrolling in TPC-C kernels to exploit instruction-level and memory-level parallelism for high-performance.

Once these optimizations are in place for a single TPC, we scale up the number of TPCs by weak scaling the ADD, SCALE, and TRIAD workloads (\fig{fig:tpc_compute_throughput}(c)). As shown, all three microbenchmarks achieve scalable improvements in compute throughput until the number of TPCs reaches between 11 and 15, eventually saturating at approximately 330, 530, and 670 GFLOPS for ADD, SCALE, and TRIAD, respectively. These values are significantly lower than Gaudi's peak compute throughput of 11 TFLOPS, which is provided by its 24 TPCs (\tab{tab:gpu_vs_hpu}). This limitation arises because all three microbenchmarks have very low operational intensity (i.e., 1/6 operations/byte for ADD, 1/4 operations/byte for SCALE, and 2/6 operations/byte for TRIAD), rendering the off-chip memory bandwidth to limit its overall performance.

Consequently, to explore how much Gaudi-2 and A100 can saturate the peak compute throughput of their vector engines (11 TFLOPS with Gaudi TPCs and 39 TFLOPS with A100 SIMD Cores),  the experiments in \fig{fig:tpc_compute_throughput}(d,e,f) artificially increase the operational intensity of ADD, SCALE, and TRIAD. Specifically, starting from the default STREAM benchmark configuration, we gradually increase the operational intensity (defined as the number of operations performed relative to the number of memory accesses) up to the point where the achieved compute throughput becomes saturated. At lower operational intensities, the three microbenchmarks become memory-bound, so Gaudi-2 demonstrates slightly higher compute throughput than A100 thanks to its $20\%$ higher memory bandwidth (\tab{tab:gpu_vs_hpu}). At higher operational intensities where the workloads become compute-bound, A100 achieves much higher compute throughput because of its $3.5\times$ higher vector computation power. In terms of compute utilization, Gaudi-2's compute throughput saturates at approximately 5.5 TFLOPS, 5.5 TFLOPS, and 10.9 TFLOPS for the ADD, SCALE, and TRIAD operations, reaching $50\%$, $50\%$, and $99\%$ of its maximum 11 TFLOPS throughput, respectively. In comparison, A100's compute throughput saturates at around 19.4 TFLOPS, 19.4 TFLOPS, and 38.2 TFLOPS for the ADD, SCALE, and TRIAD operations, similarly reaching $50\%$, $50\%$, and $98\%$ of its maximum 39 TFLOPS throughput, respectively (\mbox{\fig{fig:tpc_compute_throughput}(d,e,f)}).

\begin{myinlinebox}[Key takeaway \#2:]
{Due to the 3.5$\times$ performance gap in vector math throughput, Gaudi-2 falls short of A100 in terms of absolute non-GEMM performance. However, in terms of compute ``efficiency'', Gaudi-2 is comparable to A100 across all evaluated microbenchmarks, demonstrating the competitiveness of its design.}
\end{myinlinebox}

\subsection{Primitive ``Memory'' Operations}
\label{sect:microbenchmark_memory}

\begin{figure}[t]
\centering
\includegraphics[width=0.485\textwidth]{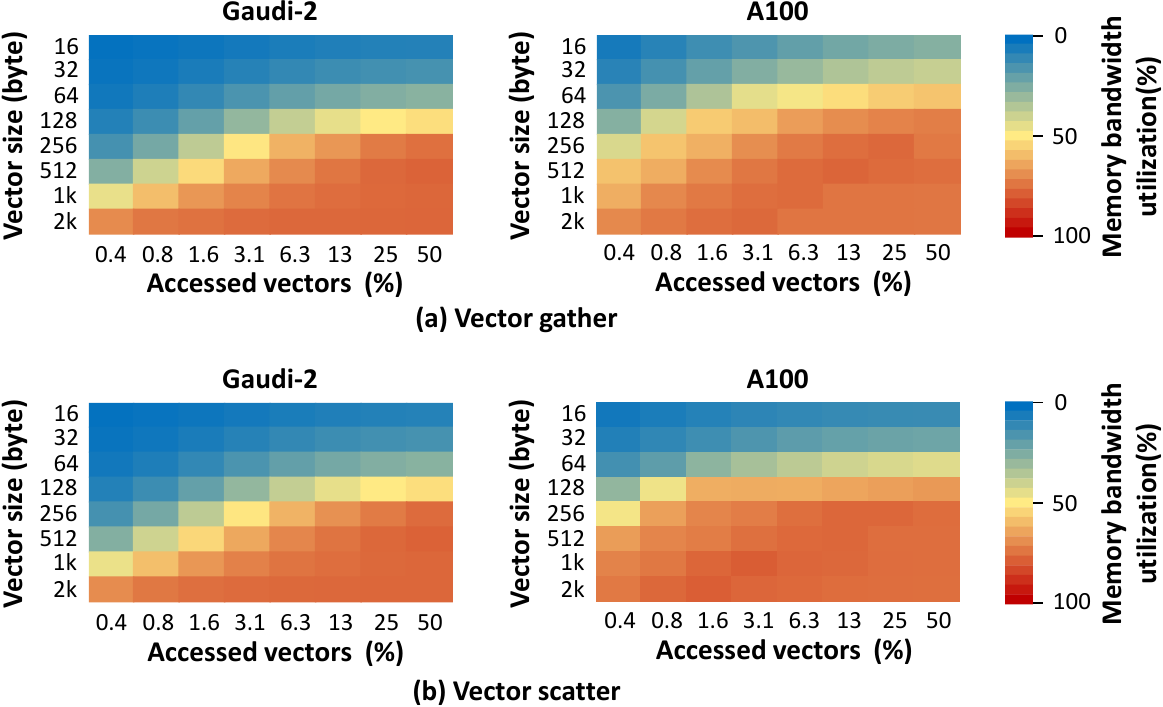} 
\caption{Memory bandwidth utilization of (a) vector gather and (b) scatter operations. The x-axis shows the proportion of vectors accessed among the total 4M vectors, i.e., the fraction of vectors either gathered from or scattered to random memory locations.}
\label{fig:mem_bw_utility_vector_gather_scatter}
\vspace{-1.4em}
\end{figure}

Our characterization of non-GEMM primitive compute operations, using the STREAM benchmark suite, confirmed the competitiveness of the Gaudi-2 memory system in handling \emph{streaming} memory access patterns. Another key aspect of memory system performance is its ability to manage \emph{random} memory accesses. Inspired by the design philosophy behind the GUPS (Giga Updates Per Second~\cite{gups}) benchmark suite, we developed our microbenchmarks to measure the memory system's performance in handling vector gather-scatter operations. These operations involve reading (vector gather) and writing (vector scatter) large amounts of data at random memory locations, which is highly memory-intensive and exhibits low data locality. The performance of these vector gather-scatter operations is particularly relevant for AI workloads like RecSys and LLMs, which require frequent embedding table lookups~\cite{tensordimm, recnmp, fafnir, rag, sasrec}.

To this end, our microbenchmarks perform vector gathers from and, similarly, vector scatters to random locations within a 2D vector array. This 2D vector array consists of 4 million vectors, with vector sizes ranging from 16 bytes to 2,048 bytes.  In \fig{fig:mem_bw_utility_vector_gather_scatter}, we show the memory bandwidth utilization of Gaudi-2 and A100 during vector gather-scatter operations. In general, Gaudi-2 achieves competitive memory bandwidth utilization when the vector size is $\ge$256 bytes, which is its minimum memory access granularity. For instance, Gaudi-2 achieves on average $64\%$ memory bandwidth utilization for $\ge$256 bytes vector gather operations, which is only slightly lower than A100's $72\%$ average memory bandwidth utilization. However, for vector sizes smaller than 256 bytes, Gaudi-2 exhibits a significant drop in memory bandwidth utility, achieving only an average $15\%$ memory throughput for $\le$128 bytes vector gather vs. A100's average $36\%$ memory bandwidth utilization, a $2.4\times$ drop in memory performance.

We speculate that the primary reason for A100's superior vector gather-scatter performance is as follows. Several prior studies focusing on reverse-engineering NVIDIA GPU microarchitecture~\cite{gpu_sector_cache_dissecting_volta, gpu_sector_cache_dissecting_memory} observed that its last-level cache either uses a cache line size of 32 bytes or a 32-byte sectored cache, suggesting that the minimum data access granularity for off-chip memory is also optimized for 32-byte data transfers~\cite{lamar, buddy_compression, nvidia_a100_gtc}. This design enables NVIDIA GPUs to fetch 32, 64, and 128 bytes from off-chip memory with minimal memory bandwidth waste, unlike Gaudi-2, which inevitably wastes bandwidth for data transfer sizes smaller than 256 bytes.

\begin{myinlinebox}[Key takeaway \#3:]
{Gaudi-2 provides competitive memory performance for regular data transfers with streaming access patterns. However, for random memory accesses like vector gather-scatter, Gaudi-2's performance falls short of A100’s when the data transfer size is smaller than its 256-byte minimum access granularity.}
\end{myinlinebox}

\subsection{Primitive ``Communication'' Operations}
\label{sect:microbenchmark_comm}

\begin{figure}[t]
\centering
\includegraphics[width=0.485\textwidth]{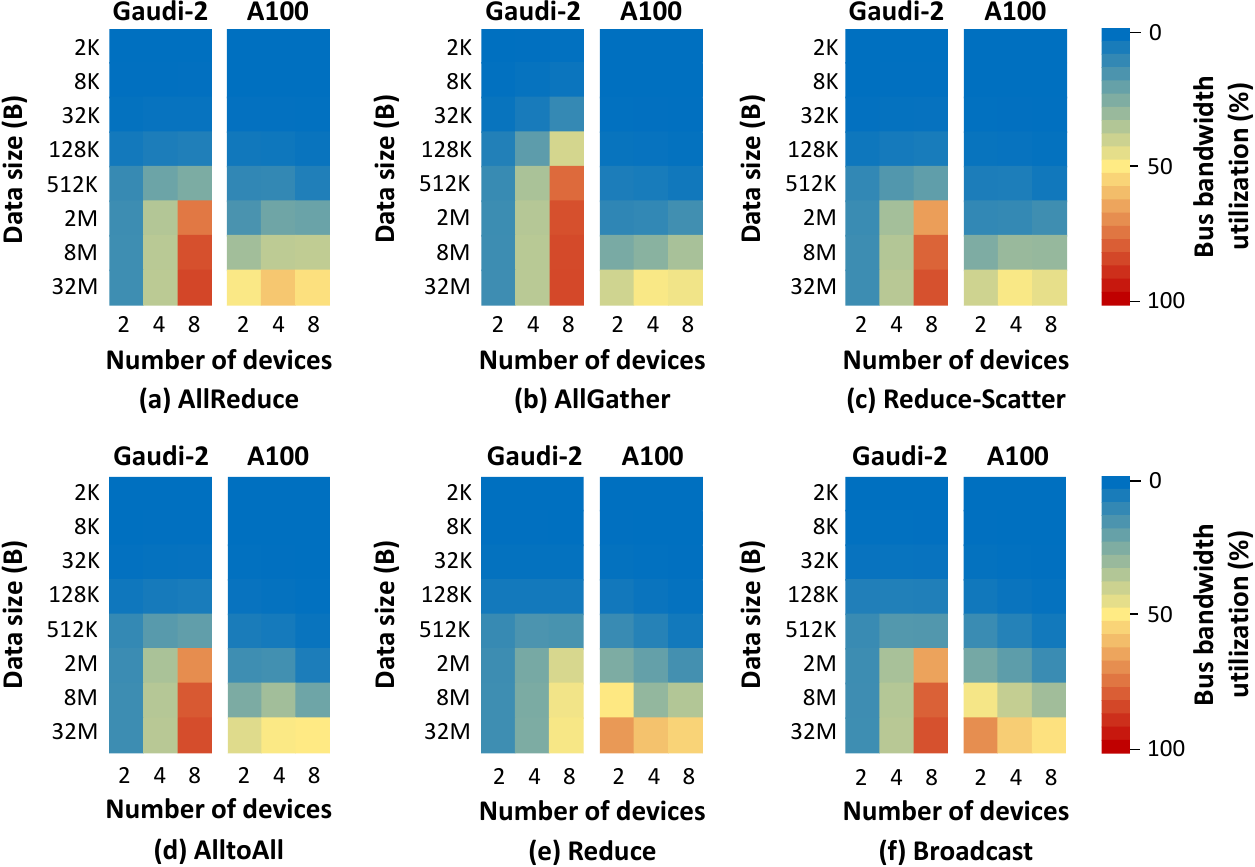} 
\caption{Bus bandwidth utilization of Gaudi-2 and A100 for collective communication operations for data sizes ranging from 2 KB to 32 MB: (a) AllReduce, (b) AllGather, (c) Reduce-Scatter, (d) AlltoAll, (e) Reduce and (f) Broadcast operations. }
\label{fig:bus_bw_utility}
\vspace{-1.0em}
\end{figure}

Recent large-scale AI models, like RecSys and LLMs, require multiple GPU or NPU devices for model serving, which necessitates frequent collective communications, such as AllReduce and Reduce-Scatter. In this subsection, we use the collective communication libraries developed by Intel and NVIDIA (HCCL~\cite{hccl} and NCCL~\cite{nccl}, respectively) to characterize the performance of six representative collective communication operations. Both Intel's HLS-Gaudi-2 and NVIDIA's DGX A100 server nodes provide an aggregate of 300 GB/sec of intra-node communication bandwidth. In \fig{fig:bus_bw_utility}, we use the bus bandwidth utilization suggested by NCCL~\cite{nccl_test} to compare the communication performance of both systems as the number of participating devices varies from 2 to 8 devices. When all eight devices participate in communication, Gaudi-2 shows higher bus bandwidth utilization than A100 for 5 of the 6 collective communication patterns evaluated. However, as the number of communicating devices decreases, Gaudi-2 experiences an almost linear decline in bus bandwidth utilization, unlike A100, whose bus bandwidth utilization remains relatively stable regardless of the number of communicating devices. As discussed in \sect{sect:gaudi_npu_hw_arch}, NVIDIA's DGX A100 server is equipped with an all-to-all network switch (NVSwitch) that enables all GPUs within the server node to communicate simultaneously, leveraging the full aggregate intra-node NVLink bandwidth. In contrast, Intel's HLS-Gaudi-2 server directly connects each pair of Gaudi-2 devices using P2P links, so the effective collective communication bandwidth scales proportionally with the number of devices involved. This setup explains the gradual decrease in Gaudi-2's bus bandwidth utilization as the number of devices used for collective communication decreases.

\begin{myinlinebox}[Key takeaway \#4:]
{The system-level collective communication performance of Gaudi-2 based system falls short of A100, not because of the limitations of the Gaudi-2 processor architecture itself, but because of its lack of an all-to-all network switch provisioned in NVIDIA's DGX A100 system. This switch enables A100 GPUs to more flexibly exploit intra-node network bandwidth, regardless of the number of devices involved in communication, a feature that is currently missing in Intel Gaudi-2 systems.}
\end{myinlinebox}

\subsection{End-to-End Application-level Analysis}
\label{sect:end_to_end_analysis}

{\bf (Models and backend software)} We now evaluate Gaudi-2 and A100 at the end-to-end AI application level, focusing on RecSys and LLMs. RecSys incorporates a heterogeneous mix of sparse and dense layers, including frontend embedding layers (which perform \emph{embedding lookups} where multiple embedding vectors are ``gathered'' from embedding tables) and backend MLP layers. Consequently, we evaluate two RecSys model configurations based on DLRM-DCNv2~\cite{dlrm_dcnv2} from the latest MLPerf benchmark suite~\cite{mlperf_recsys_benchmark} (\tab{tab:AI_workloads_table}): the compute-intensive RM1, where feature interaction and bottom/top MLP layers are dominant, and the memory-intensive RM2, where embedding layers are dominant. Because Intel Gaudi SDK currently lacks support for multi-device RecSys serving (a feature that is natively supported in TorchRec~\cite{torchrec} for serving RecSys over multi-GPUs), we focus on single-device RecSys serving for Gaudi-2. 

As for LLMs, we evaluate Llama-3.1-8B-Instruct and Llama-3.1-70B-Instruct for single- and multi-device serving~\cite{llama3}. Both systems can employ KV caches and FlashAttention~\cite{flashattn-2} by using NVIDIA's TensorRT-LLM~\cite{tensorrt-llm} for A100 and Intel's optimum-habana~\cite{optimum-habana} for Gaudi-2 as their backend engines. Finally, a synthetic dataset with an input token length fixed at 100 and output token lengths swept from 25 to 400 were used to examine the effect of fixed input-output lengths on performance. Dynamic LLM model serving scenarios with variable input-output token lengths~\cite{orca,vllm} are evaluated in \sect{subsect:programmability_vllm}. NVIDIA's CUDA Graphs~\cite{cuda_graphs} and Intel's HPU Graphs~\cite{hpu_graphs} were used as a performance tuning knob whenever appropriate and we report the highest performance achieved.

{\bf (Single-device model serving)} At the time of this writing, the embedding layer implemented in Intel's Gaudi SDK~\cite{gaudi_sdk} achieved, on average, 37\% of the performance of its GPU counterpart in the RecSys configurations we explore in this work. To assess the potential of Gaudi-2’s hardware architecture without being constrained by the limitations of the Gaudi SDK’s current implementations, we designed a custom version of the embedding layer using TPC-C. This design incorporates several key performance optimization strategies employed in the GPU-optimized FBGEMM library (\sect{subsect:programmability_dlrm} details our implementation strategy), and we utilize it as Gaudi's backend kernel library for all our evaluations in end-to-end RecSys model serving.

\begin{figure}[t!]
    \centering
    \subfloat[Performance]{
        \includegraphics[width=0.485\textwidth]{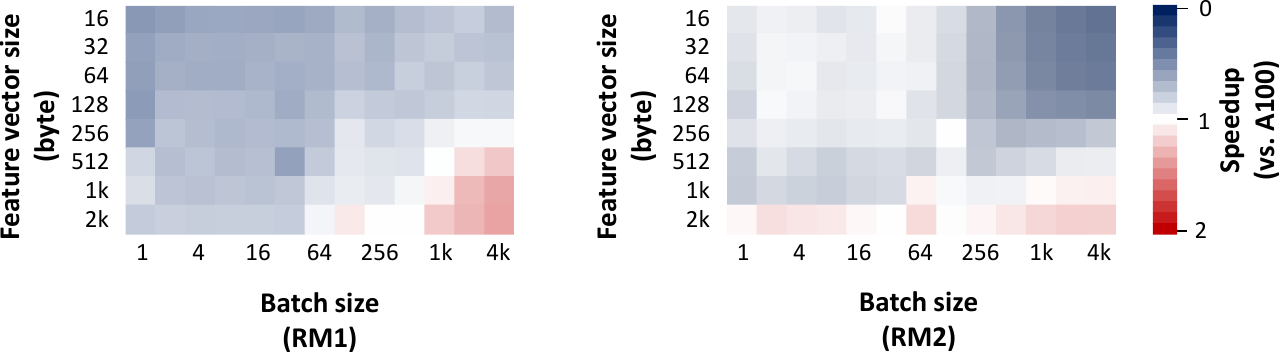} 
    }
    \vspace{0.2em}
    \centering
    \subfloat[Energy-efficiency]{
        \includegraphics[width=0.485\textwidth]{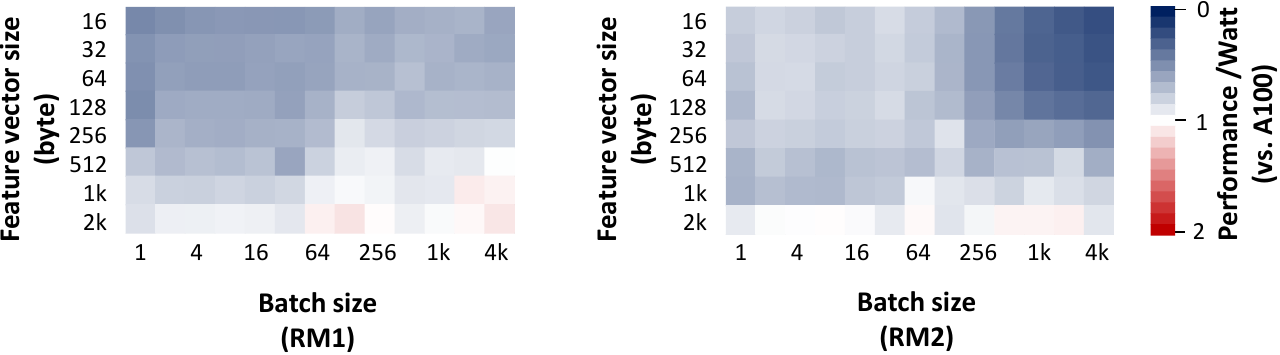} 
    }
   \vspace{-0.8em}
\caption{Gaudi-2's improvement in (a) performance and (b) energy-efficiency over  A100 when RM1, RM2 are served on a single device.
}
\label{fig:endtoendAI_perf_recsys}
\vspace{-1.5em}
\end{figure}

In \fig{fig:endtoendAI_perf_recsys}, we show Gaudi-2's end-to-end speedup (top) and energy-efficiency (bottom) over A100 for the two RecSys models, RM1 and RM2. Overall, Gaudi-2 experiences an average performance degradation of 22\% and 18\% for RM1 and RM2, respectively. While Gaudi-2 achieves higher performance with wide embedding vectors and large batch sizes (maximum 1.36$\times$ speedup), which can be attributed to Gaudi-2's higher compute throughput (\tab{tab:gpu_vs_hpu}), it generally falls short of A100 for all other model configurations. For instance, in the memory-intensive RM2, Gaudi-2 exhibits a significant performance drop in embedding vector sizes less than 256-byte, regardless of batch size (a maximum of 70\% performance loss). This is primarily due to Gaudi-2's 256-byte minimum memory access granularity and reduced memory bandwidth utilization for small number of vector gather operations, as highlighted in \sect{sect:microbenchmark_memory}. In terms of power consumption, Gaudi-2 consumed an average of 12\% more absolute power than A100 in RM1 and RM2. This result aligns with our expectations, given that Gaudi's TDP is 50\% higher than A100 (\mbox{\tab{tab:gpu_vs_hpu}}). Overall, with an increase in average latency with Gaudi-2, its end-to-end energy consumption is increased by an average 28\% than A100 for RM1 and RM2.

\begin{figure}[t!]
    \centering
    \subfloat[Performance]{
        \includegraphics[width=0.485\textwidth]{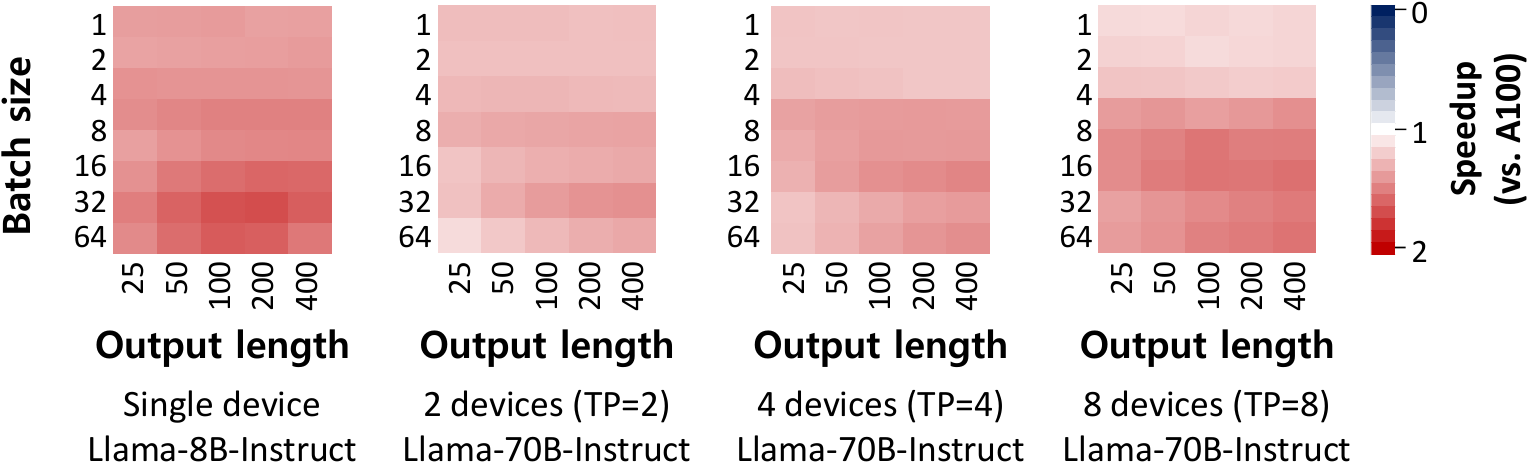} 
    }
    \vspace{0.2em}
    \subfloat[Latency breakdown]{
        \includegraphics[width=0.485\textwidth]{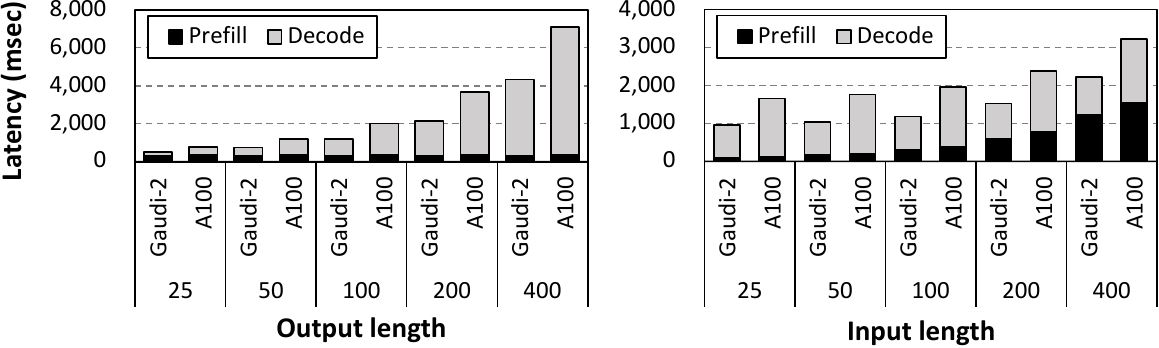} 
    }
       \vspace{-0.8em}
\caption{(a) Gaudi-2's improvement in performance over A100 when the Llama-3.1 models are served on a single device and on multiple devices. (b) For the Llama-3.1-8B-Instruct model served on a single device, we divide the latency into prefill and decoding stages, keeping the batch size fixed at 64. The left graph shows the latency breakdown when the input length is fixed at 100 while varying the output length. The right graph shows the breakdown when the output length is fixed at 100 while the input length is varied.
}
\label{fig:endtoendAI_perf_llm}
\vspace{-1.5em}
\end{figure}

\begin{figure}[t]
    \centering
    \includegraphics[width=0.485\textwidth]{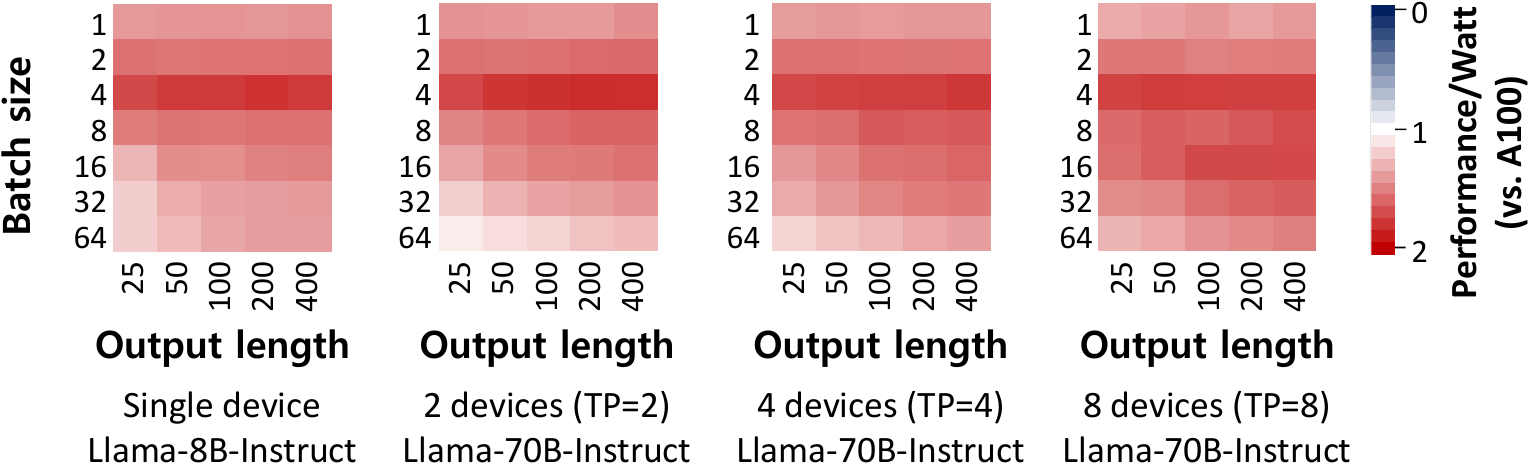} 
    \vspace{-0.8em}
\caption{Gaudi-2's improvement in energy-efficiency over A100 when the Llama-3.1 models are served on a single device and on multiple devices.}
\label{fig:endtoendAI_energy_llm}
\vspace{-0.5em}
\end{figure}

As for LLMs, the leftmost heatmap in \fig{fig:endtoendAI_perf_llm}(a) shows Gaudi-2's speedup over A100 when serving Llama-3.1-8B-Instruct over a single device. Across all batch sizes and output token lengths, Gaudi-2 consistently outperforms A100, achieving an average speedup of 1.47$\times$ (with a maximum 1.70$\times$ speedup). This performance advantage generally stems from Gaudi-2’s high peak FLOPS and memory bandwidth, benefiting both the compute-bound prefill stage and the memory-bound decoding stage (\mbox{\fig{fig:endtoendAI_perf_llm}}(b))\label{A-Q2}. Although Gaudi-2’s theoretical performance advantage over A100 is 1.4$\times$ in GEMM throughput and 1.2$\times$ in memory bandwidth (\tab{tab:gpu_vs_hpu}), Gaudi-2 achieves an even greater speedup due to its superior compute utilization across various GEMM shapes, as discussed in \sect{sect:benchmarking_compute}. The leftmost heatmap in \fig{fig:endtoendAI_energy_llm} shows a single Gaudi-2's energy-efficiency improvement over A100. In general, Gaudi-2 exhibited lower power consumption than A100 for small batch sizes, while its power consumption increased with larger batches and eventually surpassed A100 at the highest batch sizes. On average, Gaudi-2 exhibited only an average 1\% higher power consumption than A100, despite its 50\% higher TDP. As discussed in \mbox{\fig{fig:mme_graph_compiler}}(a), for small GEMM shapes, Gaudi-2 activates only a subset of its large MME units so a possible explanation for Gaudi-2's lower power consumption over small batches is that it more aggressively power-gates its circuitry via DVFS.
Overall, Gaudi-2's higher absolute performance and comparable power consumption yield an average 48\% higher energy-efficiency than A100 for single-device LLM serving. 

{\bf (Multi-device model serving)} The three rightmost heatmaps in \fig{fig:endtoendAI_perf_llm}(a)  illustrate Gaudi-2's speedup when serving the Llama-3.1-70B-Instruct model across 2, 4, and 8 devices using tensor-parallelism (TP)~\cite{megatron-LM}. The multi-device LLM serving results generally align with the trends observed in the single-device Llama-3.1-8B-Instruct serving, with average speedups of 1.29$\times$, 1.32$\times$, and 1.35$\times$ for 2, 4, and 8 devices, respectively. Interestingly, the speedup level increases with the number of devices used, attributed to the collective communication performance provided by Intel's HLS-Gaudi-2 server. As discussed in \sect{sect:microbenchmark_comm}, the use of P2P links for communication makes the performance of all-reduce (the collective communication primitive in tensor-parallelism) to be proportionally higher to the number of devices involved in collective communication, enabling higher system-level performance as more devices are employed for multi-device LLM serving. Energy-efficiency trends, illustrated in the three rightmost heatmaps in \fig{fig:endtoendAI_energy_llm}, generally align with those observed in single-device serving. Across all multi-device LLM servings, Gaudi-2 consistently demonstrates lower average power consumption, consuming around 88\% of the power of A100. Overall, Gaudi-2 achieves energy-efficiency improvements of 1.48$\times$, 1.51$\times$, and 1.56$\times$ over A100 for 2, 4, and 8 devices.

\begin{myinlinebox}[Key takeaway \#5:]
{LLM serving is primarily dominated by matrix multiplications, so Gaudi-2 demonstrates superior energy-efficiency than A100, achieving an average 48\% and 52\% improvement for single- and multi-device serving, respectively. However, RecSys relies on vector gathers and small MLP layers, so Gaudi-2 generally lags behind A100 with an average 20\% performance slowdown and an average 28\% drop in energy-efficiency.}
\end{myinlinebox}

\section{Characterizing Gaudi NPU Programmability}
\label{sect:case_study}

In this section, we present case studies on utilizing the Gaudi NPU's programming model and its software stack to conduct performance optimizations for RecSys and LLM serving systems.

\begin{figure*}[t]
  \includegraphics[width=\textwidth]{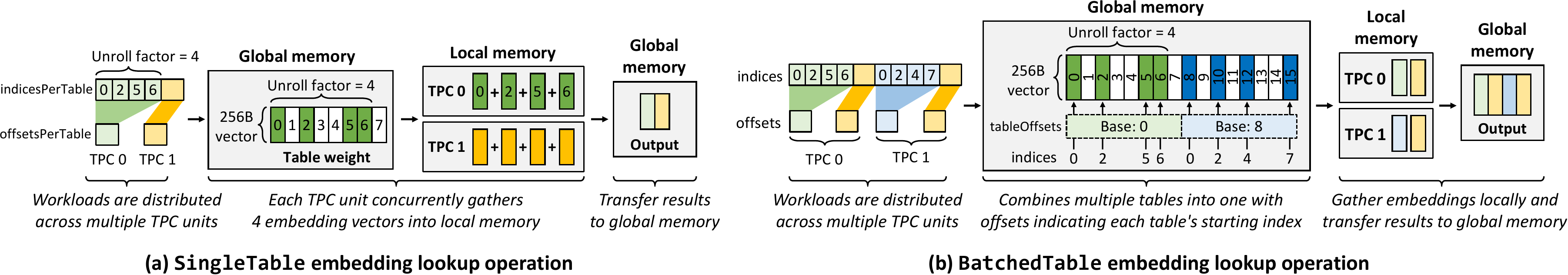} 
  \caption{Block diagrams illustrating (a) \texttt{SingleTable} embedding lookup operation, which processes embedding tables sequentially with loop unrolling to improve throughput, and (b) \texttt{BatchedTable} embedding lookup operation, which consolidates multiple tables into a single large table, enhancing memory bandwidth utilization at lower batch sizes by treating each table with offset-based indexing.}
  \label{fig:embbag_opt_diagram}
\end{figure*}

\begin{figure}[t!]
    \centering
    \includegraphics[width=0.485\textwidth]{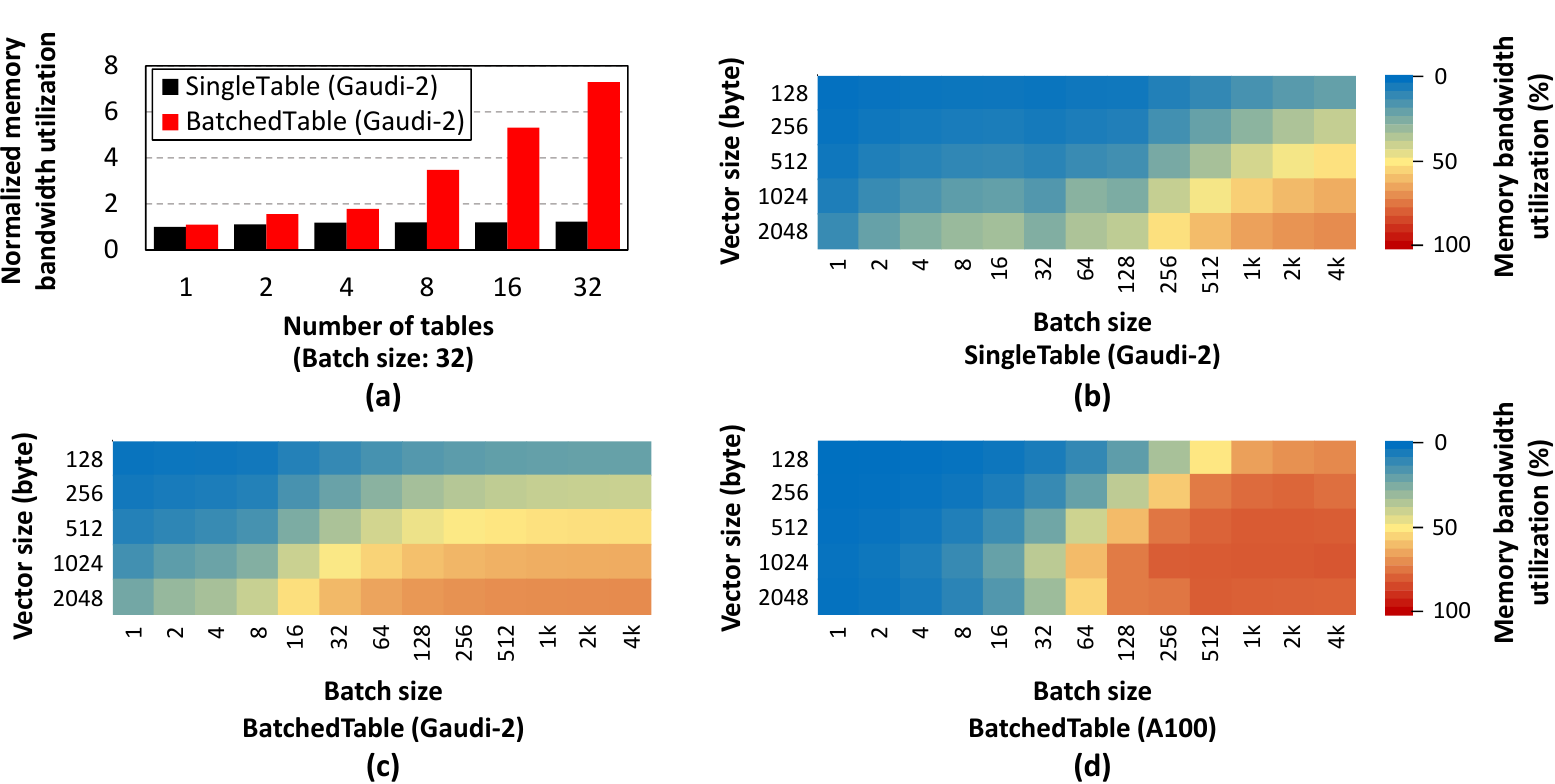}
    \caption{Memory bandwidth utilization of embedding lookup operations, using the embedding layer configuration from RM2 (\tab{tab:AI_workloads_table}). (a) Utilization is normalized to \texttt{SingleBatch} with a vector size fixed at 256 bytes while varying the numbers of tables. (b, c, d) Utilization when varying both embedding vector sizes and batch sizes.}
    \label{fig:embbag_opt_perf}
    \vspace{-1.2em}
\end{figure}

\subsection{Performance Optimization at the Low-level TPC-C: A DLRM Case Study}
\label{subsect:programmability_dlrm}
  
Meta's TorchRec library~\cite{torchrec} is built on FBGEMM’s GPU-optimized embedding lookup operator, which reduces CUDA kernel launch overhead by \emph{batching} multiple embedding tables' vector gather operations into a \emph{single} CUDA kernel execution (referred to as \texttt{BatchedTable}). Currently, Intel's Gaudi SDK does not support TorchRec, so its embedding lookup implementation does not batch vector gather operations across multiple tables. Instead, each TPC kernel launch processes only a \emph{single} table's embedding vector gathers (henceforth referred to as \texttt{SingleTable}), resulting in $N$ separate TPC kernel launches for $N$ embedding table lookups. 

To evaluate Gaudi's programmability for low-level performance optimizations, we implemented both the \texttt{SingleTable}\footnote{
As mentioned in \sect{sect:end_to_end_analysis}, the embedding lookup operator provided with Gaudi SDK (based on the \texttt{SingleTable} approach) achieved 37\% of the performance of its GPU-optimized FBGEMM counterpart. Our custom TPC-C \texttt{SingleTable} embedding lookup provides an average $60\%$ higher performance than this Gaudi SDK version.
} and the \texttt{BatchedTable} embedding lookup operators for Gaudi-2 using TPC-C. Our approach incorporates several optimizations tailored for embedding lookups as follows. The \texttt{SingleTable} operator performs embedding lookups individually for each table. The TPC-C kernel's for-loop is unrolled by a factor of 4 over embedding table lookup indices to maximize memory-level parallelism (i.e., four embedding vector gathers per each TPC are concurrently initiated for each for-loop iteration, \fig{fig:embbag_opt_diagram}(a)). The gathered embedding vectors are stored inside TPC's local memory to minimize data movement. Additionally, we distribute workloads (i.e., \texttt{offsetsPerTable} in \fig{fig:embbag_opt_diagram}(a)) across multiple TPC units to maximize chip-wide memory-level parallelism.
  
Despite these optimizations in our \texttt{SingleTable} operator, a challenge remains: with low batch sizes, a single TPC unit cannot fully utilize memory bandwidth because the workload per TPC is limited to embedding vector lookups within a \emph{single} embedding table. Even when multiple embedding tables are subject to embedding lookups, memory bandwidth remains underutilized because embedding lookups across multiple tables are performed sequentially through separate TPC-C kernel launches (i.e., memory bandwidth utilization does not increase with a larger number of tables, \fig{fig:embbag_opt_perf}(a)). To address this issue, our \texttt{BatchedTable} operator \emph{fuses} embedding lookups from multiple tables into a single TPC-C kernel. Similar to FBGEMM’s CUDA-optimized \texttt{BatchedTable}, our TPC-C \texttt{BatchedTable} implementation treats multiple tables as one large table, using a separate offset to indicate the starting index location of each table (\texttt{tableOffsets} in \fig{fig:embbag_opt_diagram}(b)). This approach requires passing indices and offsets for all tables to the TPC-C kernel in a single call. Consequently, our \texttt{BatchedTable} achieves significantly higher memory bandwidth utilization compared to \texttt{SingleTable} as the number of tables increases, as shown in \fig{fig:embbag_opt_perf}(a). It is worth noting that, with larger batch sizes, the performance gap between \texttt{SingleTable} and \texttt{BatchedTable} diminishes, as \texttt{SingleTable} can exploit more parallelism across different batches to improve memory bandwidth utilization (\fig{fig:embbag_opt_perf}(b) and (c)).

 Overall, our Gaudi-2 \texttt{BatchedTable} achieves an average memory bandwidth utilization of 34.2\% and a peak utilization of 70.5\%, representing a 1.52$\times$ improvement over \texttt{SingleTable}. In comparison, A100 demonstrates an average memory bandwidth utilization of 38.7\% with a peak of 81.8\% (\fig{fig:embbag_opt_perf}(d)). As shown in our vector gather-scatter microbenchmark experiments (\fig{fig:mem_bw_utility_vector_gather_scatter}), Gaudi-2 shows sub-optimal performance in fine-grained vector gathers so \texttt{BatchedTable} (Gaudi-2) experiences a noticeable performance drop for vector sizes below 256 bytes, with an average utilization of 12.0\%. In contrast, the A100 sustains much higher performance at these lower vector sizes, with an average utilization of 25.3\%.

\begin{figure*}[t!]
    \includegraphics[width=\textwidth]{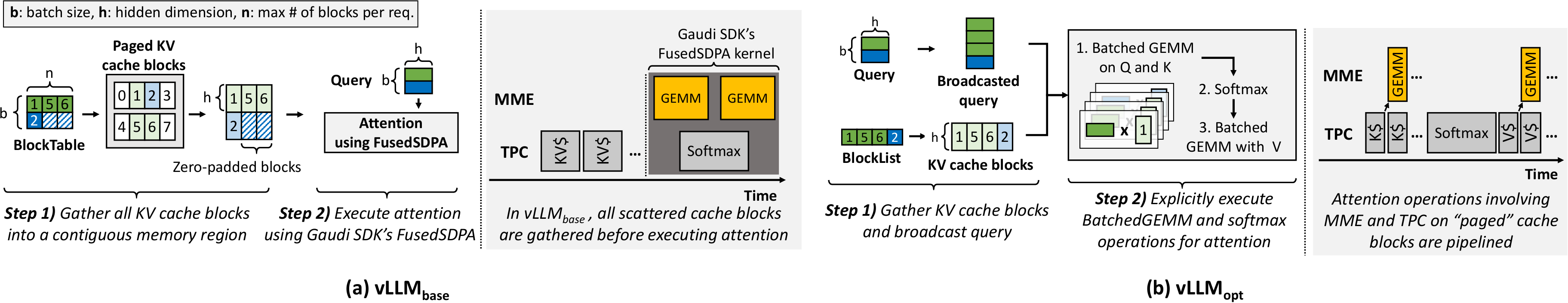} 
  \vspace{-1.5em}
  \caption{High-level overview of the PagedAttention implementation and its execution timeline in (a) baseline vLLM$_{base}$ and (b) performance-optimized vLLM$_{opt}$.}
  \label{fig:vllm_opt_diagram}
  \vspace{-0.5em}
\end{figure*}

\begin{myinlinebox}[Key takeaway \#6:]
{This case study confirmed that the TPC-C programming system provides a sufficient level of flexibility for low-level performance optimizations. Compared to state-of-the-art FBGEMM-based A100 executions, our Gaudi-2 optimized kernel for embedding layers achieved, on average, 95\% of the throughput of A100 for large embedding vector sizes ($\ge$256 bytes) but only 47\% for small vectors (<256 bytes). The noticeable performance degradation for small vectors primarily stems from A100's superior hardware architecture (which better supports fine-grained memory accesses) rather than from the differences in the programming models.}
\end{myinlinebox}

\subsection{Performance Optimization at the High-level PyTorch: A vLLM Case Study}
\label{subsect:programmability_vllm}

Serving LLMs over batched requests poses unique challenges due to the dynamic nature of input-output sequences across different requests. These variations can result in GPU memory fragmentation, which reduces the maximum batch size that the serving system can support, lowering throughput. To address this issue, vLLM~\cite{vllm} has gained widespread adoption which employs \emph{PagedAttention} that divides the key-value (KV) cache into smaller blocks, allocating them on demand rather than pre-allocating memory that would otherwise remain unused. This strategy effectively mitigates memory fragmentation, significantly increasing maximum batch size. In batched LLM serving, the attention layers~\cite{transformer} experience a significant increase in latency as batch size grows, so supporting high-performance PagedAttention is critical for vLLM.
  
While vLLM natively supports a CUDA-optimized PagedAttention kernel~\cite{vllm} for GPU-based LLM serving systems, implementing PagedAttention for Gaudi-based systems presents several unique challenges. This is because the current Gaudi SDK lacks low-level APIs that allow programmers to directly control the operation of the MME units within the user-programmed TPC-C kernel. In NVIDIA's CUDA, programmers can utilize the \texttt{WMMA} (Warp Matrix Multiply and Accumulate) APIs~\cite{wmma} to directly utilize GPU's Tensor Cores (alongside the normal CUDA Cores) for computation within the low-level CUDA kernel. However,  Gaudi programmers can only access the MME units at the PyTorch level, whose functionality is limited to the built-in, pre-compiled MME-optimized kernels provided with the Intel Gaudi SDK. Consequently, performance optimizations involving the MME units must be conducted at the PyTorch level, underscoring the role of the Gaudi graph compiler. This poses a unique challenge for implementing Gaudi-optimized PagedAttention, as it demands efficient coordination of GEMM operations and KV cache management for high performance. This contrasts with our DLRM case study in \sect{subsect:programmability_dlrm}, where the primary performance optimization target was the embedding lookup operator, which consists of vector operations that can take advantage of the programmable TPC vector unit through custom low-level kernel implementations. In this case study, we discuss performance optimization strategies that can be employed at the PyTorch level for implementing Gaudi-optimized PagedAttention. Specifically, we discuss Gaudi's PyTorch level programmability and its interaction with the Gaudi graph compiler to manage low-level hardware behavior to maximize LLM serving throughput.

\fig{fig:vllm_opt_diagram}(a) illustrates the baseline implementation of PagedAttention mechanism in Gaudi vLLM fork~\cite{vllm-fork} (hereafter referred to as \texttt{vLLM$_{base}$}). This approach uses a 2D tensor, \texttt{BlockTable}, to store the indices of KV cache blocks required by each query. When multiple requests within a single batch have varying sequence lengths, \texttt{BlockTable} is padded with zeros for queries with shorter sequence lengths (e.g., the blue-colored query in \fig{fig:vllm_opt_diagram}(a)), leading to unnecessary gathering of KV cache blocks by the TPC units. This redundant gathering of KV cache blocks, caused by zero-padded indices in \texttt{BlockTable}, results in inefficient utilization of Gaudi's compute and memory resources. Furthermore, recall that efficient use of Gaudi requires the TPC and MME to operate in parallel to hide latency, necessitating pipelined execution of the TPC-based KV cache block gather operations and MME-based GEMM operations. vLLM$_{base}$ initially gathers the scattered KV cache blocks into a contiguous memory region, and execute the \texttt{FusedSDPA}~\cite{FusedSDPA} kernel, which is a Gaudi-optimized implementation of FlashAttention~\cite{flashattn-2} (the functionality of which is equivalent to PyTorch's \texttt{scaled\_dot\_product\_attention}~\cite{pytorch_sdpa}). We observe that such implementation is not optimized for the block-based operative nature of PagedAttention as it prevents the Gaudi graph compiler from effectively pipelining its operation across MME and TPC, resulting in further performance degradation.

To alleviate the impact of redundant KV cache block gathers, a performance-optimized vLLM (vLLM$_{opt}$) replaces the 2D \texttt{BlockTable} with a 1D tensor named \texttt{BlockList} (\fig{fig:vllm_opt_diagram}(b)). By concatenating only the \emph{effectual} KV cache block indices of each request into this \texttt{BlockList}, gathers caused by zero-padding are eliminated, fetching only the KV cache blocks needed for each query. Additionally, by restructuring the query tensor shape to align with the \texttt{BlockList}-based KV cache blocks, vLLM$_{opt}$ performs \texttt{batched GEMM} across the gathered KV cache blocks, followed by the corresponding Softmax operation and others. With this approach adopted in vLLM$_{opt}$, we observe that the graph compiler more effectively partitions the TPC-based KV cache block gather operations and the MME-based batched GEMM operations into independent sub-operation slices, enabling efficient pipelined execution across TPC and MME and significantly improving hardware utilization.

\begin{figure*}[t!]
    \centering
    \includegraphics[width=\textwidth]{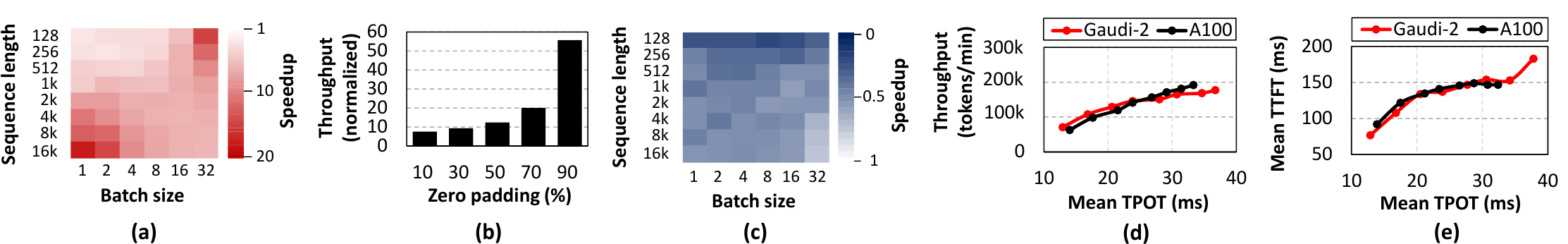} 
  \caption{(a and b) Effectiveness of vLLM$_{opt}$ on improving the performance of PagedAttention vs. vLLM$_{base}$ (results are normalized to vLLM$_{base}$):  (a) we vary the input sequence length and batch size and measure output token generation latency (fraction of zero-padded indices are 0\% in this experiment), and (b) under the sequence length=4K and batch size=32 datapoint in (a), we vary the proportion of zero-padded indices in \texttt{BlockTable} from 10 to 90\% to evaluate the effect of redundant KV cache block gathers. In (c), we show how much PagedAttention's throughput improves with vLLM$_{opt}$ by comparing it to the A100, with results normalized to the A100. In (d) and (e), we sweep the maximum decode stage batch size~\mbox{\cite{orca}} and present (d) the changes in end-to-end serving throughput and (e) the observed mean TTFT (Time-To-First-Token) and mean TPOT (Time-Per-Output-Token) values. The results are collected on a single vLLM$_{opt}$ based Gaudi-2 and A100. To properly reflect LLM serving system's dynamism and variable output length, we used the Dynamic-Sonnet dataset~\cite{sonnet_dataset}.
  }
    \label{fig:vllm_attention_perf}
\end{figure*}
  
\fig{fig:vllm_attention_perf} shows vLLM$_{opt}$'s effect in improving PagedAttention and end-to-end LLM performance. On average, vLLM$_{opt}$ achieves 7.4$\times$ improvement in PagedAttention throughput over vLLM$_{base}$ when the fraction of zero-padded indices is 0\% (\fig{fig:vllm_attention_perf}(a)). This result highlights the efficacy of vLLM$_{opt}$'s PyTorch-level performance optimizations and graph compiler's effectiveness in pipelining PagedAttention's operations across MME and TPC. Furthermore, the experiment in \fig{fig:vllm_attention_perf}(b) shows that PagedAttention's throughput improves by up to 55.7$\times$ (average 21$\times$) as the fraction of zero-padded indices increases, emphasizing the importance of eliminating redundant operations. Despite these improvements, vLLM$_{opt}$ still falls short of A100, achieving an average of 45\% of A100's PagedAttention throughput (\mbox{\fig{fig:vllm_attention_perf}}(c)). However, due to Amdahl's law and Gaudi-2's performance gains in GEMM operations in MLP layers  (\sect{sect:benchmarking_compute}), the vLLM$_{opt}$-based Gaudi-2 demonstrates similar end-to-end performance \mbox{(\fig{fig:vllm_attention_perf}(d))} to A100, with comparable sensitivity to SLO (service level objective) oriented metrics when sweeping the inference server's maximum batch size (i.e., the change in TTFT (Time-To-First-Token) vs. TPOT (Time-Per-Output-Token)  \mbox{(\fig{fig:vllm_attention_perf}(e))}). 

\begin{myinlinebox}[Key takeaway \#7:]
{This case study showed that, while Gaudi SDK's current lack of support for directly programming MMEs within the low-level TPC-C kernel imposes restrictions on programmer flexibility, graph compiler can still effectively capture the appropriate level of parallelism to better utilize its compute resources when programmed properly at the PyTorch level. Consequently, while the performance-optimized Gaudi vLLM achieved 45\% of the performance of a GPU-optimized vLLM, Gaudi-2's end-to-end LLM performance was shown to be competitive to A100.
}
\end{myinlinebox}

\section{Discussion and Future Work}
\label{sect:discussion}

{\bf (Discussion)} \sect{subsect:programmability_vllm} discussed the programming challenges associated with the black-box nature of Gaudi SDK, particularly its lack of low-level APIs for directly programming the MME units. While our vLLM case study demonstrated that PyTorch-level programming, combined with graph compiler-optimized operator scheduling, can achieve end-to-end LLM performance competitive with GPUs, there still exists a 2.2$\times$ gap in PagedAttention's performance vs. the GPU-optimized version and the absence of control and programming interfaces for Gaudi's hardware resources posed challenges in fully understanding the logic behind graph compiler's optimization passes. For instance, Gaudi's reliance on Intel's proprietary graph compiler, coupled with the lack of a direct programming interface to the MMEs, creates challenges for implementing low-level optimizations such as the kernel fusion techniques used in FlashAttention~\cite{flashattn-2}. That said, Gaudi's approach of raising the level of programming abstraction to simplify the development of high-performance AI kernels is in line with recent industry trends. For example, OpenAI's Triton~\cite{triton} also employs a Python-based programming model that abstracts many low-level GPU programming details, streamlining developer experience by handing over much of the performance optimizations to the OpenAI compiler stack. Overall, our experience with Gaudi so far suggests that the applicability of this emerging NPU device would be greatly improved by better support for low-level programming interfaces to key compute engines like MMEs, as well as more thorough documentation of Gaudi's graph compiler functionality and optimization passes.

{\bf (Future Work)} Because NVIDIA GPUs are the de facto standard in AI systems, we focused on comparing Gaudi against NVIDIA GPUs, leaving out the comparison with AMD GPUs~\cite{amd_mi300} or other NPUs~\cite{google_tpuv4,inferentia}. Possible future work includes evaluating Gaudi against these alternative platforms. Additionally, Intel claims that Gaudi NPUs are competitive to NVIDIA GPUs for training large-scale AI models requiring hundreds to thousands of devices. Analyzing Gaudi’s competitive edge against NVIDIA GPUs in training scenarios is part of our immediate future work.
Furthermore, AMD's recently announced Strix Halo (Ryzen AI Max) processor~\mbox{~\cite{amd_strix_halo}} integrates Zen  CPU cores, RDNA GPU architecture, and XDNA 2 NPU into a single SoC, offering a distinct alternative to both NVIDIA GPUs and Intel Gaudi NPUs. Unlike Gaudi, which is designed primarily for large-scale distributed AI workloads, Strix Halo targets efficient on-device AI inference with its XDNA 2 NPU, while also providing a powerful integrated GPU for mixed AI and graphics workloads. Future work could explore how Strix Halo’s unified memory architecture and heterogeneous compute capabilities compare to Gaudi NPUs and NVIDIA GPUs for AI serving.

\vspace{-0.5em}
\section{Related Work}

Emani et al.\cite{related_work_2} and Zhang et al.\cite{related_work_1} analyzed Gaudi NPU's performance with an emphasis on LLMs. These prior works lack a detailed comparative analysis against GPUs, especially from a computer architect's perspective, nor do they provide comparison from an energy efficiency standpoint. To the best of our knowledge, this work is the first and most comprehensive characterization of Gaudi NPU across multiple dimensions, using microbenchmarking, end-to-end energy analysis, and importantly, programming case studies to characterize its performance as well as programmability.

\vspace{-0.5em}
\section{Conclusion}
\label{sect:conclusion}

This paper evaluates Intel Gaudi NPUs as an alternative to NVIDIA GPUs, concluding that Gaudi NPUs have  potential to become a strong contender to NVIDIA GPUs for AI model serving. AI practitioners utilize high-level frameworks like PyTorch for model development. Our analysis suggests that, as long as Intel properly supports AI frameworks with performance-optimized backend libraries, the CUDA programming system itself might not be as formidable a ``moat''. However, we emphasize that our current assessment should not be interpreted as an overly optimistic outlook on Gaudi NPUs. NVIDIA's dominance in AI remains robust due to its comprehensive software ecosystem and we believe that Gaudi would benefit from better supporting low-level programming interfaces that facilitate more flexible programming experiences.

\bibliographystyle{ACM-Reference-Format}
\bibliography{refs}

\end{document}